\documentclass[twocolumn]{aastex631}

\usepackage{CJKutf8}
\usepackage{amsmath}
\usepackage{enumitem}
\usepackage{booktabs,pbox}
\usepackage{bm}
\usepackage{comment}
\usepackage[caption=false]{subfig}

\newcommand\ZZU{Institute for Astrophysics, School of Physics, Zhengzhou University, Zhengzhou 450001, China}
\newcommand\HAUT{College of Science, Henan University of Technology, Zhengzhou 450000, China}
\newcommand\HUUC{School of Mathematics and Physics, Henan University of Urban Construction, Pingdingshan 467036, China}
\newcommand\NJU{School of Astronomy and Space Science, Nanjing University, Nanjing 210023, China}

\begin{document}

\title{Study of hybrid stars with nonstrange quark matter cores}

\author[0000-0002-9159-8129]{Cheng-Ming Li}\thanks{E-mail: licm@zzu.edu.cn}
\affiliation{\ZZU}

\author{He-Rui Zheng}
\affiliation{\ZZU}

\author{Shu-Yu Zuo}
\affiliation{\HAUT}

\author[0000-0003-3351-9565]{Ya-Peng Zhao}\thanks{E-mail: zhaoyapeng2013@hotmail.com}
\affiliation{\HUUC}

\author[0000-0001-7379-2975]{Fei Wang}
\affiliation{\ZZU}

\author[0000-0001-7199-2906]{Yong-Feng Huang}
\affiliation{\NJU}

\begin{abstract}
In this work, under the hypothesis that quark matter may not
be strange \citep{PhysRevLett.120.222001}, we adopt
a modification of the coupling constant of the
four-quark scalar interaction
$G\rightarrow G_1+G_2\langle\bar{\psi}\psi\rangle$
in the 2-flavor Nambu-Jona-Lasinio (NJL) model to study
nonstrange hybrid stars,
where $G_1$ and $G_2$ are two parameters
constrained by using the lattice QCD simulation results
at the critical temperature and zero chemical potential.
The Maxwell construction is used to describe the first-order
confinement-deconfinement phase transition in hybrid stars.
With recent measurements on neutron star mass, radius,
and tidal deformability, the hybrid equation of states
are constrained. It is found that pure nonstrange
quark matter cores can exist in hybrid stars,
possessing $0.026-0.04$ solar mass. The maximum hybrid
star mass in the framework of the modified NJL model is
about 0.1 solar mass lighter than that in the
conventional 2-flavor NJL model.
It is argued that the binary neutron stars
in GW170817 should be hadron stars.
\end{abstract}


\section{Introduction}
\label{sec:intro}

The binary neutron star (BNS) merger GW170817
opened a new era of multi-messenger astronomy
~\citep{PhysRevLett.119.161101,2041-8205-848-2-L12,
2041-8205-850-2-L19,2041-8205-850-2-L34,PhysRevD.96.123012,
PhysRevLett.120.172703,PhysRevLett.120.172702,PhysRevD.97.084038,
PhysRevD.97.083015,PhysRevD.97.021501,2041-8205-852-2-L29,
2041-8205-852-2-L25,0004-637X-857-1-12,2018ApJ...862...98Z,
2018ApJ...860...57A,ma2019pseudoconformal,PhysRevD.101.043003,
PhysRevD.101.063023,Miao_2021,PhysRevLett.128.161101,Zou_2022,
PhysRevD.106.116009,PhysRevD.108.063002}. More and more
astronomical observations on neutron stars arise,
facilitating the study of neutron star structure and
equation of state (EOS). As natural laboratories to
investigate the dense strongly interacting matter,
neutron stars have been attracting much attention
in astrophysics and theoretical physics. In general,
the characteristic temperature of neutron stars can be
well described by zero temperature approximation,
due to their excessively high energy density in the interior,
thus the quantum chromodynamics (QCD) needs to be employed
to study the EOS in neutron stars. It is believed that
the density in the core of neutron stars could reach
5-10 $\rho_0$, where $\rho_0=0.16$ fm$^{-3}$ is the nuclear
saturation density~\citep{doi:10.1126/science.1090720,
annurev:/content/journals/10.1146/annurev-nucl-102419-124827}.
As a result, the hadron-quark phase transition is very likely
to happen and the deconfined quark matter will appear.
In this case, neutron stars are essentially hybrid stars.
However, it is difficult to give a unified description of the
hadronic matter, quark matter and the hadron-quark
phase transition with a single theoretical framework.
Thus the hadronic matter and quark matter in hybrid stars
are separately described with different EOSs at present,
and a certain construction scheme needs to be employed
to combine them to get a complete EOS.

As we know, the results of different effective models can be
quantitatively or even qualitatively different. Even for the
same model, if different modifications are taken into account,
the results can also be different. For example, Fig.~10 of
\citet{doi:10.1146/annurev-astro-081915-023322} shows that
the EOSs given by different effective models are different
from each other, and the corresponding mass-radius ($M-R$)
relations of neutron stars are also different. Thus there is
not a definite answer to the EOS of dense strongly interacting
matter at zero temperature at present.

To describe the hadronic matter in hybrid stars,
the EOS developed by Akmal, Pandharipande \& Ravenhall
(APR) with $A18+\delta\nu+UIX^{\ast}$ interaction is
employed in this work~\citep{PhysRevC.58.1804},
in which the Argonne $\nu_{18}$ two-nucleon interaction
and boost corrections to the two-nucleon interaction
as well as the three-nucleon interaction are
taken into account. The hadronic matter in the context of
the APR model is one kind of
charge-neutral and beta-stable fluid whose pressure and
baryon chemical potential are equilibrated.
However, to describe the quark matter in hybrid stars,
the lattice QCD is confronted with difficulties at
low-temperature and high-density regions because of the
``sign problem'', thus we need to use effective models,
such as the Nambu-Jona-Lasinio (NJL) model
~\citep{RevModPhys.64.649,Buballa2005205,li2019kurtosis,
sym13081410,PhysRevD.105.094015,sym15020541},
which manifests the spontaneous breaking of chiral symmetry.

In the framework of the NJL-type model, many studies focused
on hybrid stars~\citep{ayriyan2021bayesian,universe6060081,
alvarez2016new,PhysRevC.105.035802}, aiming to explain the
observed two-solar-mass (2 $M_{\odot}$) compact stars.
In \citet{ayriyan2021bayesian,universe6060081,alvarez2016new},
the 2-flavor NJL-type model was used to consider
the scalar quark-antiquark interaction, anti-triplet scalar
diquark interactions and vector quark-antiquark interactions
~\citep{ayriyan2021bayesian},
a chemical potential dependence of the vector mean-field
coupling $\eta(\mu)$ and a chemical potential-dependent bag
constant $B(\mu)$ ~\citep{universe6060081},
multiquark (4- and 8-quark) interactions
~\citep{alvarez2016new}, respectively.
In \citet{PhysRevC.105.035802}, the 3-flavor SU(3) NJL model
was adopted with the four-quark scalar, vector-isoscalar
and vector-isovector interactions as well as the
't Hooft interaction. Different from the above studies,
we adopt a modification of the coupling constant
of four-quark scalar interactions as
$G\rightarrow G_1+G_2\langle\bar{\psi}\psi\rangle$,
which can be regarded as a representation of an effective
gluon propagator (See Sec.~\ref{one} for specific analysis).

As for the hadron-quark phase transition,
the most widely used approach is the Maxwell construction
~\citep{10.1143/PTP.115.337,PhysRevD.80.125014,PhysRevD.80.123009},
assuming that the first-order phase transition
occurs~\citep{PhysRevD.46.1274,Bhattacharyya_2010} and
stable quark matter cores exist in hybrid stars.
However, many studies showed that hybrid stars are unstable
against oscillations in this case, because star masses
decrease with the increase of the central density,
thus quark matter cores may not exist in neutron-star interiors
~\citep{ozel2006soft,PhysRevLett.117.032501,PhysRevD.107.103009}.
In \citet{ozel2006soft} and \citet{PhysRevLett.117.032501},
the theoretical modeling of bursting neutron-star spectra
and top-down holographic model for strongly interacting
quark matter were employed, respectively, to demonstrate that
the 2 $M_{\odot}$ neutron star has ruled out soft EOSs of
neutron-star matter, and no quark matter exists in
massive neutron stars. Recently, it has been argued that
as the density increases, the boundaries of hadrons disappear
gradually and the corresponding phase transition is a crossover
~\citep{Baym_2018}. According to this assumption,
the three-window modeling~\citep{0004-637X-764-1-12,
Masuda01072013} in the crossover region was proposed.
Many studies has constructed hybrid EOSs in this scheme,
and the corresponding maximum masses of hybrid stars
are compatible with 2 $M_{\odot}$
~\citep{PhysRevD.91.045003,PhysRevD.95.056018,PhysRevD.97.103013,
PhysRevD.98.083013,Li_2022,PhysRevD.107.103009}.

In addition to theoretical studies of hadron-quark
phase transitions and hybrid EOSs, astronomical observations
of neutron star masses, radii, and tidal deformability
have also placed constraints on numerous EOSs.
Some massive neutron star observations such as
PSR J0348+0432~\citep{Antoniadis1233232} and
PSR J0740+6620~\citep{cromartie2020relativistic} require
EOSs should not be too soft, but the tidal deformability
constrained in BNS merger event GW170817 indicates
the EOSs should not be too stiff~\citep{PhysRevLett.119.161101,
PhysRevLett.121.161101}. Recently, the joint
$M-R$ observations of neutron stars from NASA's
Neutron Star Interior Composition Explorer (NICER) missions
have also imposed some constraints on these EOSs
~\citep{Riley_2019,Miller_2019,Riley_2021,Miller_2021}.
In \citet{PhysRevLett.122.061102,Miao_2020},
the authors claim that the gravitational-wave (GW) emission
of GW170817 supports a first-order hadron-quark
phase transition at supranuclear densities.

In this work, inspired by a recent work that the quark matter
may not be strange~\citep{PhysRevLett.120.222001}, we will
study nonstrange hybrid EOSs and hybrid stars with the
Maxwell construction. The hadronic EOS and quark EOS are
described by the APR model
and a modified 2-flavor NJL model, respectively.
The parameter space of $G_1$ and $G_2$ in the
modified NJL model will be fixed according to the
lattice results at zero chemical potential
~\citep{annurev:/content/journals/10.1146/annurev.nucl.53.041002.110609}.
With recent measurements on neutron star
mass, radii, and tidal deformability, the hybrid EOSs
will be constrained to get the parameter space of the
current quark mass.
To ensure that hybrid stars are stable
against oscillations, maximum masses of hybrid stars and
the masses of their quark matter cores are determined.

It is known that the Bayesian analysis is a good approach
to constrain the EOSs. Researchers have obtained
important information about the EOS of QCD
in this way~\citep{ayriyan2021bayesian,universe6060081,
alvarez2016new,PhysRevC.105.035802,alvarez2020studying,
universe5020061,ayriyan2015new,Miller_2020}. However,
considering that the lagrangian of the NJL model is convenient
for numerical calculation, in this work we have performed
calculations focusing on the EOS to get the corresponding
hybrid star $M-R$ relations and tidal deformability,
and then compared them with the relevant neutron star
astronomical observations. The model parameters and
properties of hybrid stars are constrained as well.

This paper is organized as follows. In Sec.\ref{one},
the modified NJL model for nonstrange quark matter is briefly
introduced, and the EOSs of quark matter are derived.
In Sec.\ref{two}, the Maxwell construction is used
to get hybrid EOSs. With recent astronomical observations of
neutron star mass, radius, and tidal deformability,
we constrain the hybrid EOSs. For comparison,
the $M-R$ relations and tidal deformability results of
hybrid stars from six representative hybrid EOSs are presented.
Finally, a brief summary is given in Sec.\ref{three}.

\section{EOS of nonstrange quark matter} \label{one}

As an effective model to describe cold dense quark matter,
the NJL model~\citep{RevModPhys.64.649,Buballa2005205} is
widely used in the study of hybrid stars and quark stars.
    In this work, we consider a modified
    version, in which the Lagrangian has the following form:
\begin{equation}\label{genLag}
  \mathcal{L}=\bar{\psi}(i{\not\!\partial}-m)\psi+
    (G_1+G_2\langle\bar{\psi}\psi\rangle)
  [(\bar{\psi}\psi)^2+(\bar{\psi}i\gamma^5\tau\psi)^2],
\end{equation}
where $m$ is the current quark mass (because of an exact
isospin symmetry between $u$ and $d$ quarks adopted in
this work, $m_{\rm u}=m_{\rm d}=m$).
    Different from the normal NJL model, we
    adopt $G_1+G_2\langle\bar{\psi}\psi\rangle$ to represent
    the four-fermion coupling strength, where
    $\langle\bar{\psi}\psi\rangle$ is the quark condensate.
    The term $(G_1+G_2\langle\bar{\psi}\psi\rangle)
    [(\bar{\psi}\psi)^2+(\bar{\psi}i\gamma^5\tau\psi)^2]$
describes interactions in scalar-isoscalar and
pseudoscalar-isovector channels.

    In the following we will clarify the
    modification in detail.
Based on our current knowledge of strong interactions, the
coupling constant $G$ in the normal
NJL model can be regarded as a representation of an effective
gluon propagator. In light of QCD theory, the quark and gluon
propagators should satisfy their respective Dyson-Schwinger
(DS) equations, and these two equations are coupled with
each other. It is demonstrated that quark propagators in the
Nambu phase and Wigner phase are very different from each other
~\citep{Cui2018,0954-3899-45-10-105001,PhysRevD.99.076006},
so it can be inferred that the corresponding gluon propagators
in these two phases are also different~\citep{hong2006quark}.
However, in the normal NJL model, $G$ is simplified as a
constant, remaining the same in these two phases. In addition,
according to simulations of lattice QCD, the gluon propagator
should vary with temperature, although its dependence on the
chemical potential is still uncertain. In the normal NJL model,
as a representation of an effective gluon propagator,
the coupling constant $G$ is ``static'', and thus cannot
fulfill the requirement of lattice QCD.

In the QCD sum rule approach~\citep{REINDERS19851}, it is
argued that the full Green function can be divided into two
parts: the perturbative part and nonperturbative part. The
condensates can be expressed as various moments of
nonperturbative Green function. As a result, the most general
form of the ``nonperturbative'' gluon propagator is
\begin{equation}\label{gapeq}
  D_{\mu\nu}^{\rm{npert}}\equiv
  D_{\mu\nu}^{\rm{full}}-D_{\mu\nu}^{\rm{pert}}\equiv
  c_1\langle\bar{\psi}\psi\rangle+c_2\langle
  G^{\mu\nu}G_{\mu\nu}\rangle+...,
\end{equation}
where $\langle G^{\mu\nu}G_{\mu\nu}\rangle$ refers to the
gluon condensate, $c_1$ and $c_2$ are coefficients which can
be calculated in the QCD sum rule approach
~\citep{Steele1989,pascual1984qcd}, and the ellipsis represents
the contributions from other condensates, such as the mixed
quark-gluon condensate. Among these condensates, the quark
condensate possesses the lowest dimension, and a nonzero value
of it, in the chiral limit, precisely signifies the dynamical
chiral symmetry breaking. Therefore, it plays the most
important role in the QCD sum rule approach.
In this work, we will deal with its contribution separately,
and the contribution of other condensates is simplified into
the perturbative part of the gluon propagator.
In the normal NJL model, it is equivalent to a modification
of the coupling constant $G$ in the following way
~\citep{PhysRevD.85.034031,Cui2013,Cui2014,PhysRevD.93.036006,PhysRevD.94.096003,doi:10.1142/S0217732317501073,Fan_2019,PhysRevD.97.103013},
\begin{equation}\label{effG}
  G\rightarrow G_1+G_2\langle\bar{\psi}\psi\rangle.
\end{equation}

Now the coupling strength $G$ will depend on both $u$ and $d$
quark condensates via this modification, where $G_2$ refers to
the weight factor of the influence of the quark propagator on
the gluon propagator. It should be noted that following
this modification,
    we need to correspondingly change the term
    $G\langle\bar{\psi}\psi\rangle^2$ in the mean-field
    thermodynamic potential of the normal NJL model to ensure
    the consistency in thermodynamics.
Similar to \citet{PhysRevD.52.5206}, the term
$G\langle\bar{\psi}\psi\rangle^2$ is
replaced by the integral over $\langle\bar{\psi}\psi\rangle$ of
$2\langle\bar{\psi}\psi\rangle{\rm
d}(G\langle\bar{\psi}\psi\rangle)/{\rm d}
\langle\bar{\psi}\psi\rangle$, in which d.../d... denotes
derivative, and the result is
$G_1\langle\bar{\psi}\psi\rangle^2+4G_2\langle\bar{\psi}\psi\rangle^3/3$.
In this way,
    the mean-field thermodynamic potential has
    the following form,
\begin{eqnarray}
  \Omega(T,\{\mu\},\{\langle\bar{\psi}\psi\rangle\})=
  &&\Omega_{\rm M}(T,\mu)
    +G_1\langle\bar{\psi}\psi\rangle^2
    +\frac{4}{3}G_2\langle\bar{\psi}\psi\rangle^3
  \nonumber\\
  && + {\rm const},\label{thermpot}
\end{eqnarray}
where $\Omega_{\rm M}$ denotes
the contribution of a gas of quasiparticles,
\begin{eqnarray}
  \Omega_{\rm M}=&&-2N_{\rm c}N_{\rm f}\int\frac{{\rm d}^3
  p}{(2\pi)^3}\{T\,{\rm ln}(1+{\rm
  exp}(-\frac{1}{T}(E_{p}-\mu)))\nonumber\\
  &&+T\,{\rm ln}(1+{\rm exp}(-\frac{1}{T}(E_{p}+
  \mu)))+E_{p}\},\label{thermopotofquasi}
\end{eqnarray}
    where $E_{p}=\sqrt{\overrightarrow{p}^2+M^2}$ is the quark
    on-shell energy. Note that when $G= {\rm const}$
    (i.e., $G_1=G$ and $G_2=0$), our model reproduces the
    condensate term in the normal NJL model.

The effective mass of the constituent quark is now given by
\begin{eqnarray}
  M &=& m-2
    (G_1+G_2\langle\bar{\psi}\psi\rangle)
  \langle\bar\psi\psi\rangle.\label{gap}
\end{eqnarray}
The quark condensate $\langle\bar\psi\psi\rangle$
and the particle number density $\rho$
can be derived from $\Omega$ as
\begin{eqnarray}
  \langle\bar\psi\psi\rangle
    &=& \frac{\partial\Omega}{\partial m}\nonumber \\
    &=& -2 N_{\rm c}N_{\rm f}
        \int\frac{{\rm d}^3p}{(2\pi)^3}
        \frac{M}{E_{p}}[1-n_{p}(T,\mu)-\overline{n}_{p}(T,\mu)],\nonumber\\
        \label{psibarpsi} \\
  \rho &=& -\frac{\partial \Omega}{\partial \mu}\nonumber\\
       &=& 2 N_{\rm c}N_{\rm f}
           \int\frac{{\rm d}^3 p}{(2\pi)^3}
           (n_{p}(T,\mu)-\overline{n}_{p}(T,\mu)),\label{rho}
\end{eqnarray}
where $n_{p}(T,\mu)$ and $\overline{n}_{p}(T,\mu)$ are
the Fermi occupation numbers of quarks and antiquarks,
respectively, which are defined as
\begin{eqnarray}
  n_{p}(T,\mu) &=&
  [{\rm exp}^{(E_{p}-\mu)/T}+1]^{-1},\label{fonq} \\
  \overline{n}_{p}(T,\mu) &=&
  [{\rm exp}^{(E_{p}+\mu)/T}+1]^{-1}.\label{fonaq}
\end{eqnarray}
Because the NJL model cannot be renormalized, the proper-time
regularization is adopted in the following calculations.
In addition, we need to fix the parameter set
($\Lambda_{\rm UV}$, $G$) to fit experimental data
($f_{\pi}=92$ MeV, $M_{\pi}=135$ MeV) at zero temperature and
chemical potential. The parameter fixing process is similar to
that of \cite{Cui:2014hya}.

Although the lattice QCD is confronted with
the ``sign problem'' at finite chemical potentials,
the simulating results at zero chemical potential
can still help us determine the values of $G_1$ and $G_2$.
According to the simulations of lattice QCD,
the chiral phase transition at zero chemical potential
is a crossover, and the corresponding
pseudo-critical point is located at $T_{\rm pc}=173\pm8$ MeV
in the 2-flavor case
~\citep{annurev:/content/journals/10.1146/annurev.nucl.53.041002.110609}.
Different from the meaning of the so-called ``critical point''
in the case of first-order phase transition,
the ``pseudo-critical point'' here refers to the condition
that the crossover occurs, and its position can be identified
by the peak of susceptibilities,
such as the chiral susceptibility $\chi_s$ in
\citet{annurev:/content/journals/10.1146/annurev.nucl.53.041002.110609},
which is defined as
$\chi_s=-\partial\langle\bar{\psi}\psi\rangle/\partial m$
~\citep{PhysRevD.88.114019}.

\begin{figure}
\includegraphics[width=0.47\textwidth]{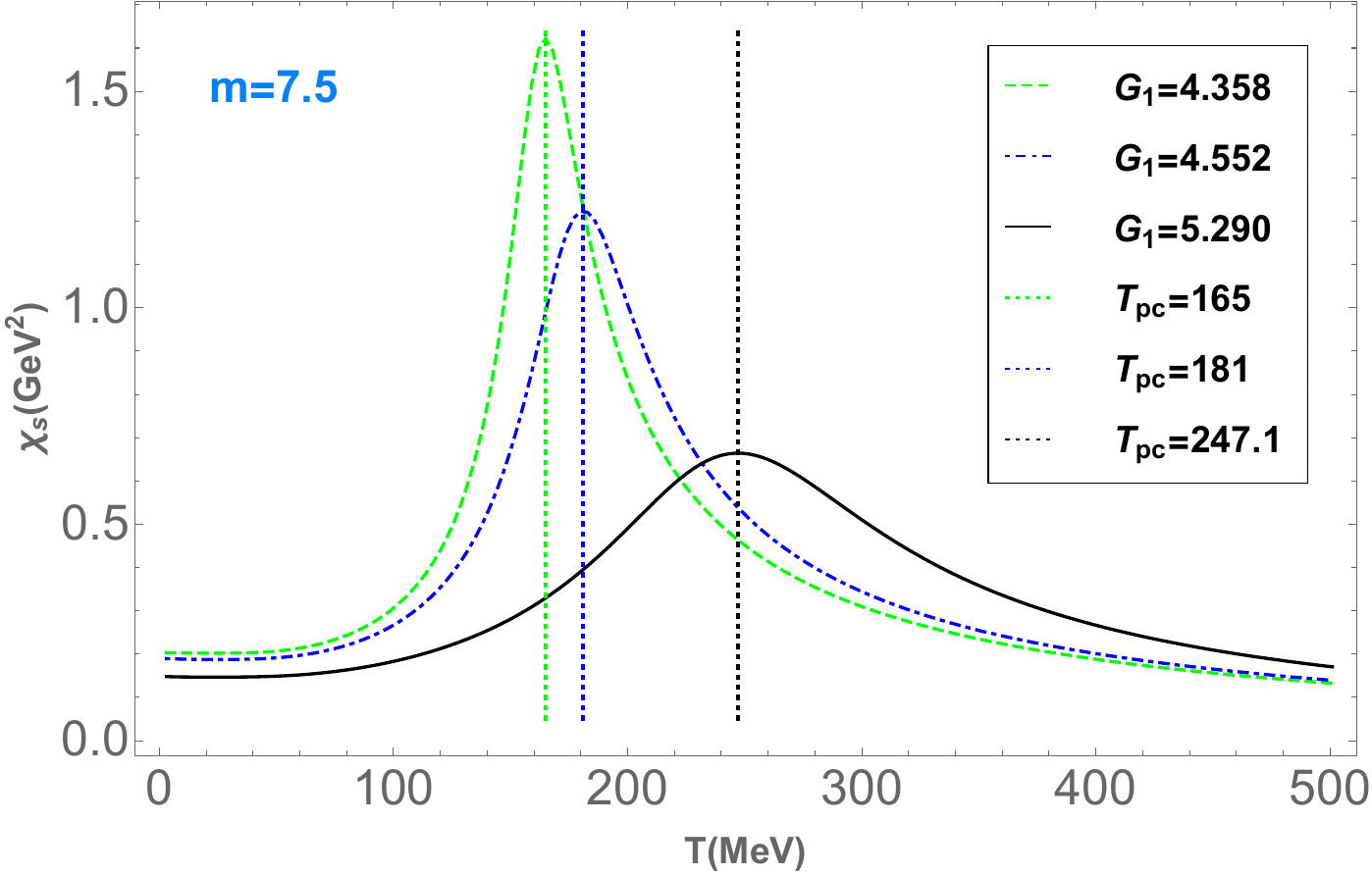}
\caption{The chiral susceptibilities $\chi_{\rm s}$
versus temperature for different $G_1$. $m$ is fixed as
7.5 MeV. The three vertical lines denote the peak place of
$\chi_{\rm s}$, i.e., $T_{\rm pc}=165, 181, 247.1$ MeV.}
\label{Figure1}
\end{figure}

\begin{figure}
\includegraphics[width=0.47\textwidth]{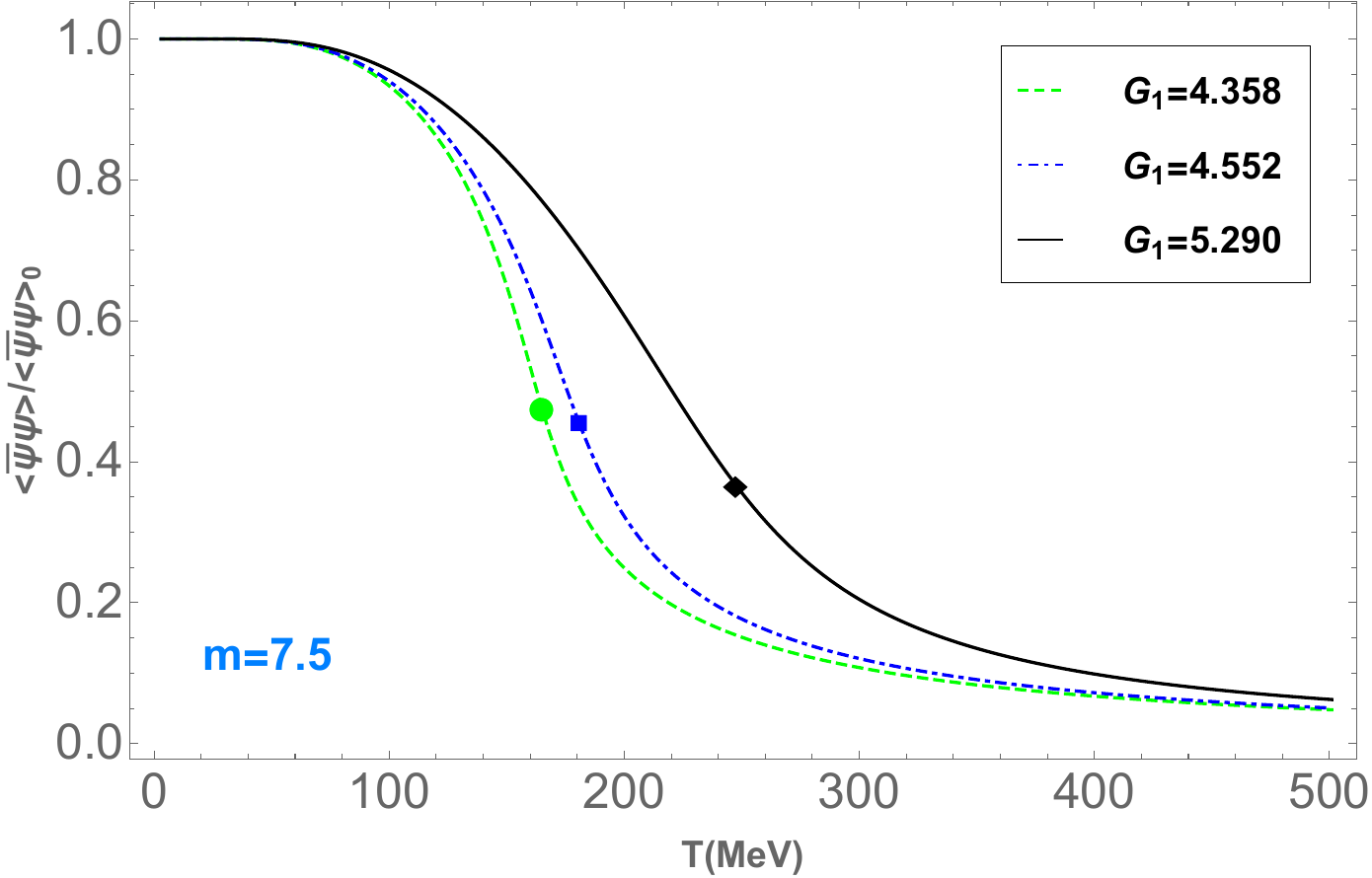}
\caption{The scaled quark condensates versus $T$
for different $G_1$. $m$ is fixed as 7.5 MeV.
The corresponding pseudo-critical points
are also marked on these curves, respectively.} \label{Figure2}
\end{figure}

According to our calculations, when the
current quark mass $m$ is in the range of 3.5 -- 10 MeV,
$T_{\rm pc}$ will be larger than 181 MeV in the
conventional NJL model. Generally, $T_{\rm pc}$ will decrease
as $G_1$ decreases. In Fig.~\ref{Figure1}, the chiral
susceptibility for different $G_1$ is plotted, taking
$\mu=0$ and $m=7.5$ MeV. As $G_1$ varies from 4.358 to
4.552 GeV$^{-2}$, $T_{\rm pc}$ changes from 165 to 181 MeV,
satisfying the constraint from lattice simulations. However,
when $G_1=G=5.29$ GeV$^{-2}$, $T_{\rm pc}$ in the conventional
NJL model is 247.1 MeV, which is much larger than the
corresponding value of lattice simulations.

The QCD sum rule at the renormalization
scale of 1 GeV suggests the $u$ quark condensate should
be $229 \pm33$ MeV$^3$ ~\citep{dosch1998direct}
or $242\pm15$ MeV$^3$ ~\citep{jamin2002flavour}.
Considering this constraint,
the whole parameter sets of the modified 2-flavor NJL model
in this work are shown in Table~\ref{Table1}.
For a certain current quark mass, we take
two boundary values for $G_1$, corresponding to the cases
with $T_{\rm pc}=165$ and 181 MeV, respectively. Similar to
previous studies under the proper-time
regularization~\citep{RevModPhys.64.649,KOHYAMA2015682,PhysRevD.95.056018},
the effective quark masses in Table~\ref{Table1} are in a range
of 190 -- 250 MeV, which are smaller than 300 MeV.
As a comparison, it has been shown that for the same current
quark mass, the three-momentum cutoff regularization leads to
a larger effective quark mass and a smaller momentum cutoff
~\citep{RevModPhys.64.649,Ratti2007,KOHYAMA2015682,Kohyama:2016fif}.

\begin{table}
\centering
\caption{Parameter sets in this work.}
\label{Table1}
\begin{tabular}{p{0.7cm}p{0.7cm}p{0.7cm}p{0.9cm}p{1.0cm}p{1.0cm}p{1.0cm}}
\hline\hline
$\quad m$&$\,\,\Lambda_{\rm UV}$&$\,\,M$&
${-}\langle\bar{u}u\rangle^{\frac{1}{3}}$&$\quad\,\,G$&
$\quad\,\,\, G_{1}$&$\quad\,\,\,G_{2}$\\
$[{\rm MeV}]$&$[{\rm MeV}]$&$[{\rm MeV}]$&
$\,\,[{\rm MeV}]$&$[{\rm GeV}^{-2}]$&$[{\rm GeV}^{-2}]$&
$[{\rm GeV}^{-5}]$\\\hline
    $\,\,\,\,$4.5&$\,$1151&$\,\,\,191$&$\,\,\,\,\,258$&$\,\,\,2.727$&$\quad2.563$&$\,\,\,-4.791$\\
    $\,\,\,\,$4.5&$\,$1151&$\,\,\,191$&$\,\,\,\,\,258$&$\,\,\,2.727$&$\quad2.627$&$\,\,\,-2.924$\\\hline
    $\,\,\,\,$5.8&$\,$1002&$\,\,\,206$&$\,\,\,\,\,237$&$\,\,\,3.761$&
    $\quad3.375$&$\,\,\,-14.50$\\
    $\,\,\,\,$5.8&$\,$1002&$\,\,\,206$&$\,\,\,\,\,237$&$\,\,\,3.761$&
    $\quad3.485$&$\,\,\,-10.36$\\
    \hline
    $\,\,\,\,$7.5&$\,$873&$\,\,\,225$&$\,\,\,\,\,217$&$\,\,\,5.290$&
    $\quad4.358$&$\,\,\,-45.29$\\
    $\,\,\,\,$7.5&$\,$873&$\,\,\,225$&$\,\,\,\,\,217$&$\,\,\,5.290$&
    $\quad4.552$&$\,\,\,-35.86$\\
    \hline
    $\,\,\,\,$9.5&$\,$774&$\,\,\,250$&$\,\,\,\,\,201$&$\,\,\,7.399$&
    $\quad5.226$&$\,\,\,-133.8$\\
    $\,\,\,\,$9.5&$\,$774&$\,\,\,250$&$\,\,\,\,\,201$&$\,\,\,7.399$&
    $\quad5.572$&$\,\,\,-112.5$\\
    \hline\hline
\end{tabular}
\end{table}

Fig.~\ref{Figure2} plots the scaled order parameter of
chiral phase transition
($\langle\bar{\psi}\psi\rangle/\langle\bar{\psi}\psi\rangle_0$)
versus temperature when $m=7.5$ MeV.
We can find that
$\langle\bar{\psi}\psi\rangle/\langle\bar{\psi}\psi\rangle_0$
decreases smoothly from one to zero as temperature increases,
thus the transition at $\mu=0$ is the crossover,
consistent with the simulation result of lattice QCD.

\begin{figure}
\includegraphics[width=0.47\textwidth]{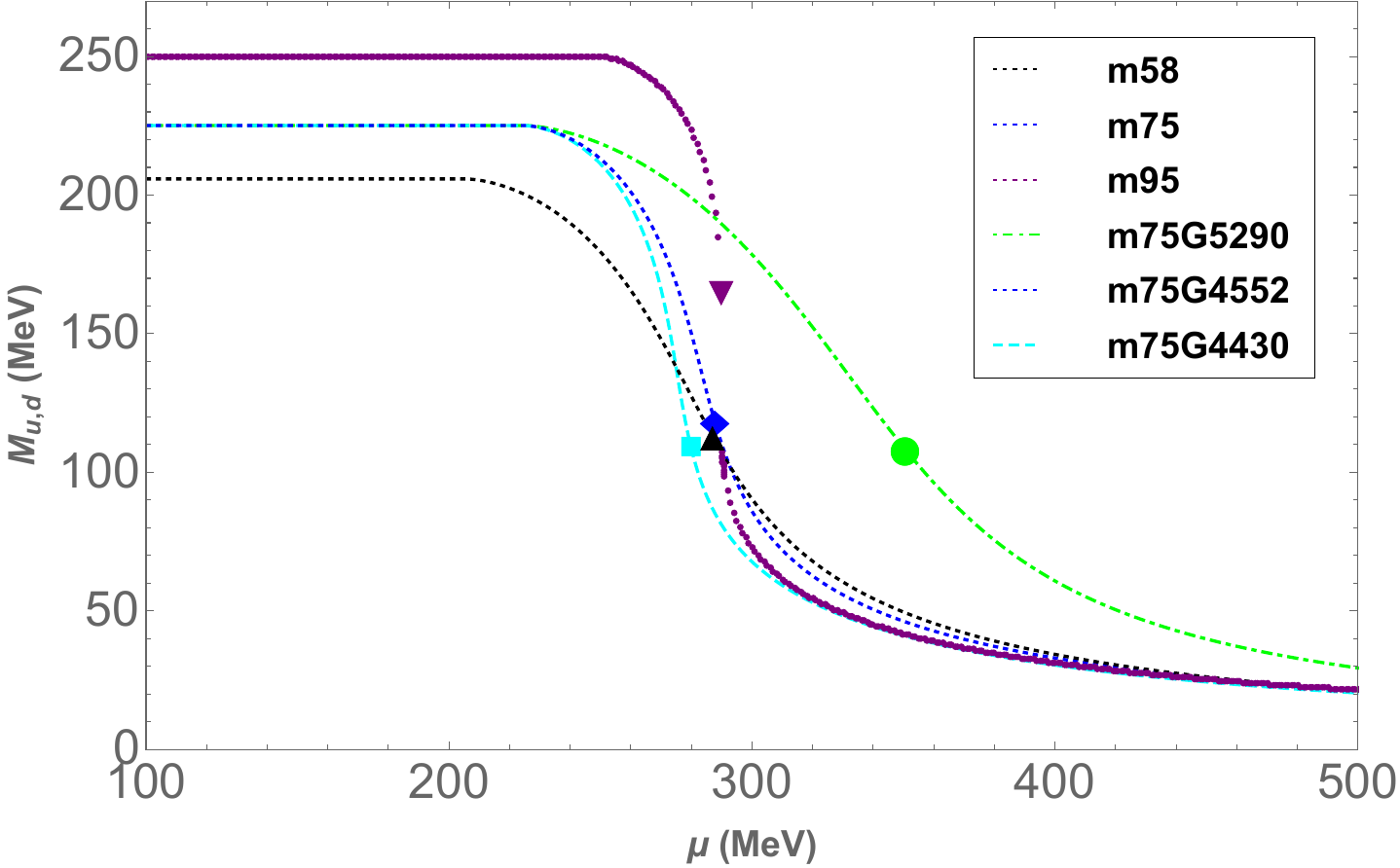}
\caption{The effective mass of quarks versus
chemical potentials $\mu$ at $T=0$.
Note that m58, m75, m95 correspond to
three cases with $m=5.8$, 7.5, 9.5 MeV, respectively,
in which we have taken $T_{\rm pc}=181$ MeV.
Similarly, in the three cases of m75G5290, m75G4552 and
m75G4430, we have taken $m=7.5$ MeV and
$G_1=5.290$, 4.552, 4.43 GeV$^{-2}$, respectively.
The (pseudo-)critical points are marked on the lines
correspondingly.}
\label{Figure3}
\end{figure}

Now we extend our calculation to finite chemical potentials
at $T=0$ to get EOSs of the quark matter. After solving
Eq.~(\ref{gap}) with the modification of Eq.~(\ref{effG}),
we can get the dependence of effective quark mass $M$
on the chemical potential, which is shown in
Fig.~\ref{Figure3}. It can be seen that
for the dense quark systems with a fixed
$T_{\rm pc}$, a larger $m$ leads to a larger effective quark
mass in the vacuum, and thus a larger gap will emerge
when the chiral phase transition occurs. Specifically,
when $T_{\rm pc}=181$ MeV, the crossover occurs for the systems
with $m<9$ MeV, and the first order phase transition occurs
for the systems with $m>9$ MeV. The critical chemical
potential ($\mu_{\rm c}$) is around 285 MeV.
As a comparison, we also present the
results for a fixed $m$ but with different values of $G_1$
in Fig.~\ref{Figure3}. As $G_1$ increases, $\mu_{\rm c}$ will
also increase. Note that the green dashed line in
Fig.~\ref{Figure3} corresponds to normal NJL model, in which
the pseudo-critical chemical potential is about
350 MeV. It is much larger than the corresponding
results of those cases with $G_1\in(4.430, 4.552)$ GeV$^{-2}$.

In the framework of the NJL model, it is demonstrated that
whether the first-order chiral phase transition occurs
at $T=0$ (when $m\neq0$) depends on the regularization scheme
that is employed
~\citep{Buballa2005205,doi:10.1142/S0217732316500863,KOHYAMA2015682}.
In \citet{Buballa2005205}, the three-momentum cutoff
regularization is used and a first-order phase transition
happens at $T=0$. However, in
\citet{doi:10.1142/S0217732316500863},
the authors use the PTR and find a crossover in the
phase transition region at $T=0$. Actually,
in \citet{KOHYAMA2015682}, it is clarified that
the low current quark mass ($m\leq4$ MeV) can result in
a crossover at $T=0$ for both PTR and three-momentum cutoff
regularization.

According to lattice QCD simulations,
we notice that at zero temperature and under the
PTR, when $m>7.5$ MeV, $\mu_{\rm c}$ of the modified NJL
model is significantly lower than that in the normal NJL
model. Sometimes even a first-order phase transition
occurs. For example, from the phase diagram of
the normal NJL model derived by \citet{KOHYAMA2015682},
we see that the crossover can only occur at $T=0$
when $m<10$ MeV. However, according to our calculations,
a first-order phase transition has already occurred
when $m>9$ MeV in the modified NJL model.

\begin{figure}
\includegraphics[width=0.47\textwidth]{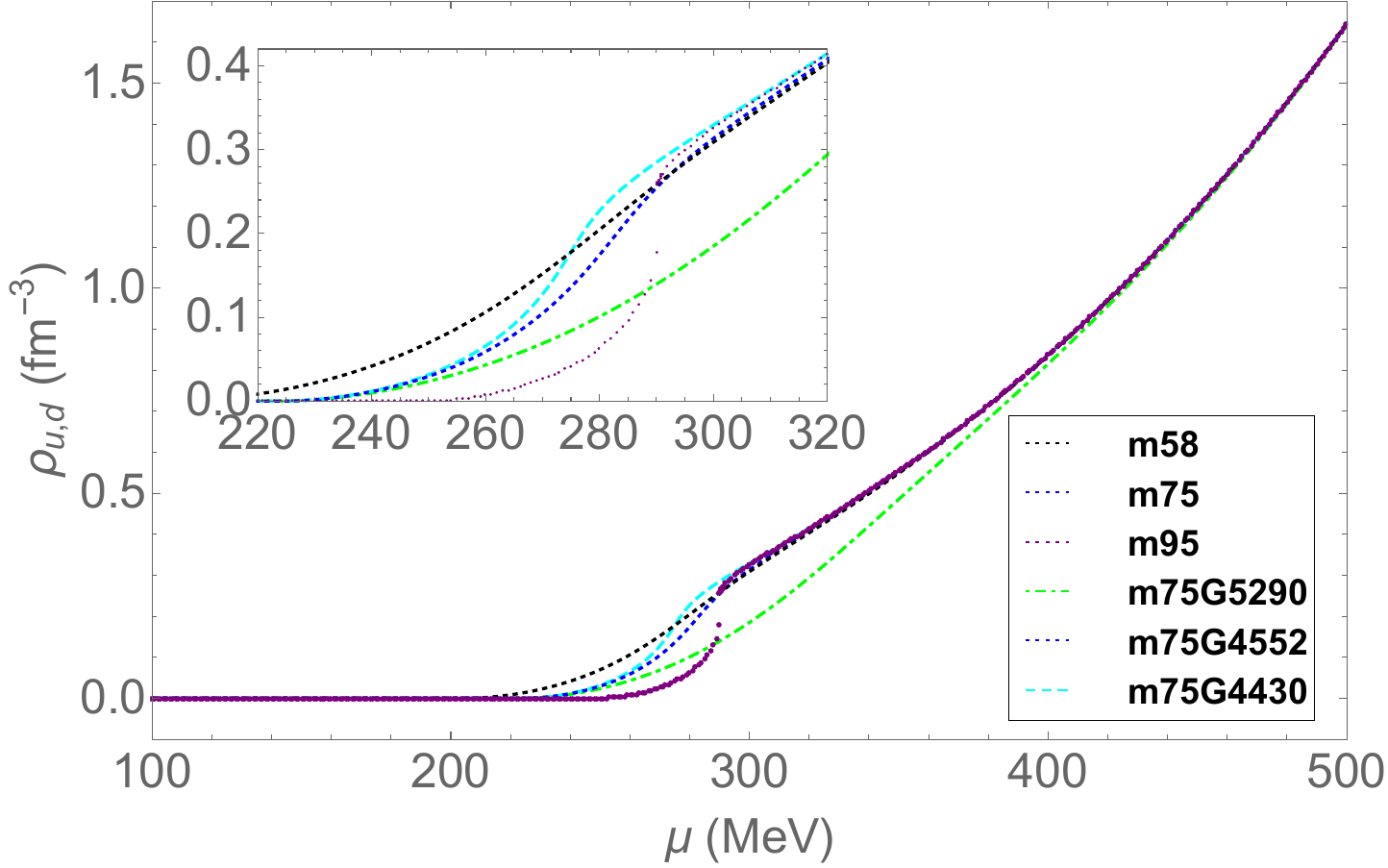}
\caption{The number density of $u, d$ quarks versus
chemical potentials $\mu$ at $T=0$.
Line styles are the same as those in Fig.~\ref{Figure3}.}
\label{Figure4}
\end{figure}

The dependence of quark number density
on the chemical potential
at $T=0$ is shown in Fig.~\ref{Figure4}. We can see that
a larger $m$ leads to a lower
quark number density when $T_{\rm pc}=181$ MeV and
$\mu\in(220-290)$ MeV. For a fixed $m$ with different
values of $G_1$, the quark number densities are
only slightly different in the crossover region according
to the lattice QCD simulation results on $T_{\rm pc}$.
However, they are quite different in normal NJL models.

\begin{figure}
\includegraphics[width=0.47\textwidth]{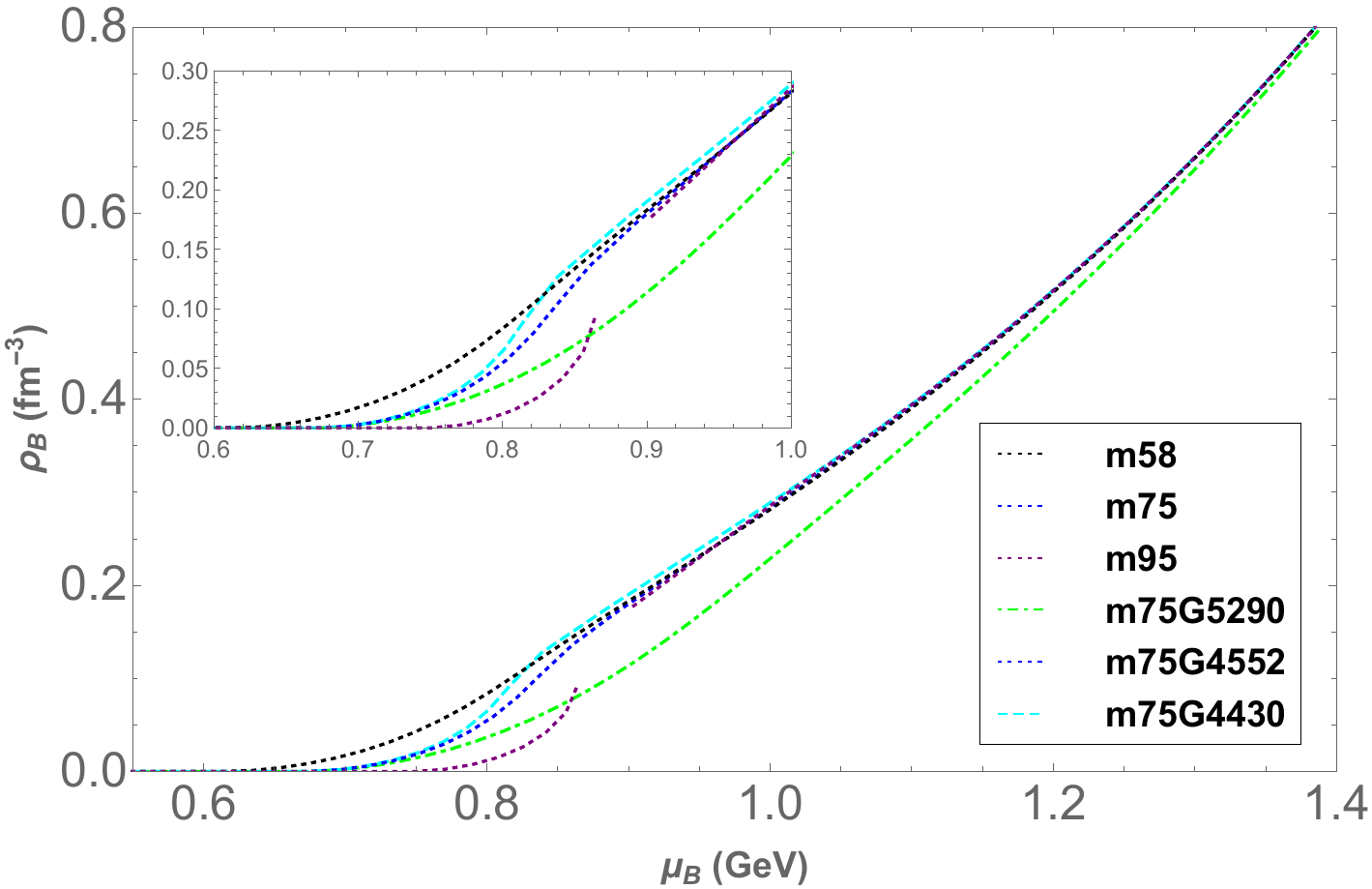}
\caption{The baryon density versus baryon chemical potential
of the quark system. Line styles are the same as those in
Fig.~\ref{Figure3}.} \label{Figure5}
\end{figure}

To describe the strongly interacting matter in hybrid stars,
we need to consider the beta equilibrium and
electric charge neutrality,
\begin{eqnarray}\label{constrains}
  &&\mu_{\rm d}=\mu_{\rm u}+\mu_{\rm e}, \nonumber \\
  &&\frac{2}{3}\rho_{\rm u}-\frac{1}{3}\rho_{\rm
  d}-\rho_{\rm e}=0,
\end{eqnarray}
where $\rho_{\rm e}=\mu_{\rm e}^3/3\pi^2$ is the number density
of electrons at $T=0$. The relation between the baryon density
$\rho_{\rm B}$ and baryon chemical potential $\mu_{\rm B}$ is
shown in Fig.~\ref{Figure5},
where $\mu_{\rm B}=\mu_{\rm u}+2\mu_{\rm d}$ and
$\rho_{\rm B}=(\rho_{\rm u}+\rho_{\rm d})/3$.
In Fig.~\ref{Figure5}, the shape of each curve is quite
similar to that in Fig.~\ref{Figure4}. Interestingly,
in the modified NJL model, the $\rho_{\rm B}-\mu_{\rm B}$
relation of the quark system is almost identical
even when different parameter sets are adopted.
The difference mainly occurs in the phase transition region.

According to the definition, the EOS of dense quark matter
at $T=0$ is~\citep{doi:10.1142/S0217751X08040457}
\begin{equation}\label{EOSofQCD}
  P(\mu)=P(\mu=0)+\int_{0}^{\mu}d\mu'\rho(\mu'),
\end{equation}
and the energy density of the quark system
can be expressed as~\citep{PhysRevD.86.114028,PhysRevD.51.1989}
\begin{equation}\label{rbedasp}
  \epsilon=-P+\sum_{i=u,d,e}\mu_{\rm i}\rho_{\rm i}.
\end{equation}
Note that $P(\mu=0)$ in Eq.~(\ref{EOSofQCD}) is irrelevant to
the chemical potential. It represents the negative vacuum
pressure, which corresponds to the vacuum bag constant ($-B$)
in the MIT bag model. In the NJL model, it can be calculated
by using Eq. (2.60) of \citet{Buballa2005205} under the
mean field approximation.
    In this way, the bag constant is regarded
    as the pressure difference between the trivial and
    the non-trivial vacuum. It is known that the trivial and
    non-trivial vacuum can be described by the (quasi-)Wigner
    and Nambu solution of the gap equation, respectively. As
suggested by
\citet{0954-3899-45-10-105001,Cui2018,PhysRevD.99.076006,PhysRevD.100.123003},
the vacuum pressure is defined as
\begin{eqnarray}\label{Bagconstant}
      P(\mu=0)&=&P(M_N)-P(M_W) \nonumber\\
              &=&\Omega(0,0,\langle\bar{\psi}\psi\rangle_W)-
                 \Omega(0,0,\langle\bar{\psi}\psi\rangle_N),
\end{eqnarray}
where $M_N$ and $M_W$ represents the Nambu and quasi-Wigner
solution of the gap equation, respectively, and
$\langle\bar{\psi}\psi\rangle_N$ and
$\langle\bar{\psi}\psi\rangle_W$
are the corresponding quark condensates.

\section{Structure of hybrid stars}
\label{two}

Under the Maxwell construction scheme, the first-order
hadron-quark phase transition occurs
when the baryon chemical potentials and pressures of
these two phases are equal,
\begin{equation}\label{EOSofQCD2}
  P_{\rm H}(\mu_{\rm B,c})=P_{\rm Q}(\mu_{\rm B,c}),
\end{equation}
where $\mu_{\rm B,c}$ is the critical baryon chemical potential
of the hadron-quark phase transition,
which is around 1.5 and 1.6 GeV in the modified NJL model
and the normal NJL model, respectively.
Note that the dense hadronic matter in this work is described
by the APR EOS with $A18+\delta\nu+UIX^{\ast}$
interaction~\citep{PhysRevC.58.1804},
which is strongly favored by recent neutron star observations.
The hybrid EOS can be written as
\begin{equation}\label{habridEOS}
  P(\mu_{\rm B})=\left\{\begin{array}{lcl}
           P_{\rm H},\,\,\,when\,\, \mu_{\rm
           B}\leq\mu_{\rm B,c}, \\
           P_{\rm Q},\,\,\,when\,\, \mu_{\rm
           B}\geq\mu_{\rm B,c}.
         \end{array}\right.
\end{equation}
The corresponding energy density of the hybrid EOS is
\begin{equation}\label{habridenergy}
  \epsilon(\mu_{\rm B})=\left\{\begin{array}{lcl}
           \epsilon_{\rm H},\,\,\,when\,\, \mu_{\rm
           B}\leq\mu_{\rm B,c}, \\
           \epsilon_{\rm Q},\,\,\,when\,\, \mu_{\rm
           B}\geq\mu_{\rm B,c},
         \end{array}\right.
\end{equation}
where $\epsilon_{\rm Q}$ is the energy density of
the quark system.

In Sec.~\ref{one}, the parameter sets have been preliminarily
constrained by lattice simulations, and the results are shown
in Table~\ref{Table1}. Here they will be further constrained
based on recent measurements on neutron star mass, radius, and
tidal deformability. According to our calculations,
a 2 $M_{\odot}$ neutron star can provide a strong constraint
on the lower limit of $G_1$. It rules out the parameter sets
with $m<5.8$ MeV and $m>9.5$ MeV. On one hand,
when $m<5.8$ MeV, the most massive hybrid star mass cannot
reach 2 $M_{\odot}$. On the other hand, when $m>9.5$ MeV,
the pressure of quark matter is higher than that of
hadronic matter, and the most massive quark star is also
lighter than 2 $M_{\odot}$.

In the following, to study the influence of various parameters
on the EOS and on the structure of hybrid stars,
we choose five representative quark EOSs and their
corresponding hybrid EOSs for comparison.
The corresponding parameter sets are
$m=5.8$ MeV with $G_1=3.485$ GeV$^{-2}$, $m=7.5$ MeV
with $G_1=4.430, 4.552, 5.290$ GeV$^{-2}$, respectively,
and $m=9.5$ MeV with $G_1=5.572$ GeV$^{-2}$. Specifically,
the parameter sets with $(m, G_1)=$(5.8 MeV, 3.485 GeV$^{-2}$),
(7.5 MeV, 4.552 GeV$^{-2}$), (9.5 MeV, 5.572 GeV$^{-2}$)
correspond to the case of $T_{\rm pc}=181$ MeV, from which
we can see the influence of $m$ on the results. The parameter
sets with $(m, G_1)=$(7.5 MeV, 4.430 GeV$^{-2}$), (7.5 MeV,
4.552 GeV$^{-2}$), (7.5 MeV, 5.290 GeV$^{-2}$) correspond to
the case of $m=7.5$ MeV, from which we can find the influence
of $G_1$ on the results.

According to the discussion on the vacuum
pressure $P(\mu=0)$ in Eq.~\ref{Bagconstant},
the bag constant $B^{1/4}$ corresponding to
the parameter sets of
$(m, G_1)$=(5.8 MeV, 3.485 GeV$^{-2}$),
(7.5 MeV, 4.430 GeV$^{-2}$), (7.5 MeV, 4.552 GeV$^{-2}$),
(9.5 MeV, 5.572 GeV$^{-2}$) can be calculated as
122.7 MeV, 120.9 MeV, 123.6 MeV, 125.4 MeV,
respectively. Comparing with the normal NJL model with
$(m, G_1)$ = (7.5 MeV, 5.290 GeV$^{-2}$)
and $B^{1/4}=137.1$ MeV, the modified NJL model
under the constraint of $T_{\rm pc}$ has a smaller $B^{1/4}$,
which ranges in 120 -- 125 MeV.
For a fixed $T_{\rm pc}$, a larger $m$ will lead to a
slightly larger $B^{1/4}$.
Note that the range of $B$ in this work is consistent
with that in ~\citet{PhysRevD.46.3211}.
Considering the medium effect, the study
~\citep{LU1998443} also indicated the decline of $B$
in the framework of the quark-meson coupling model.

\begin{figure*}
\centering
\subfloat{\includegraphics[width=9cm,height=6cm]{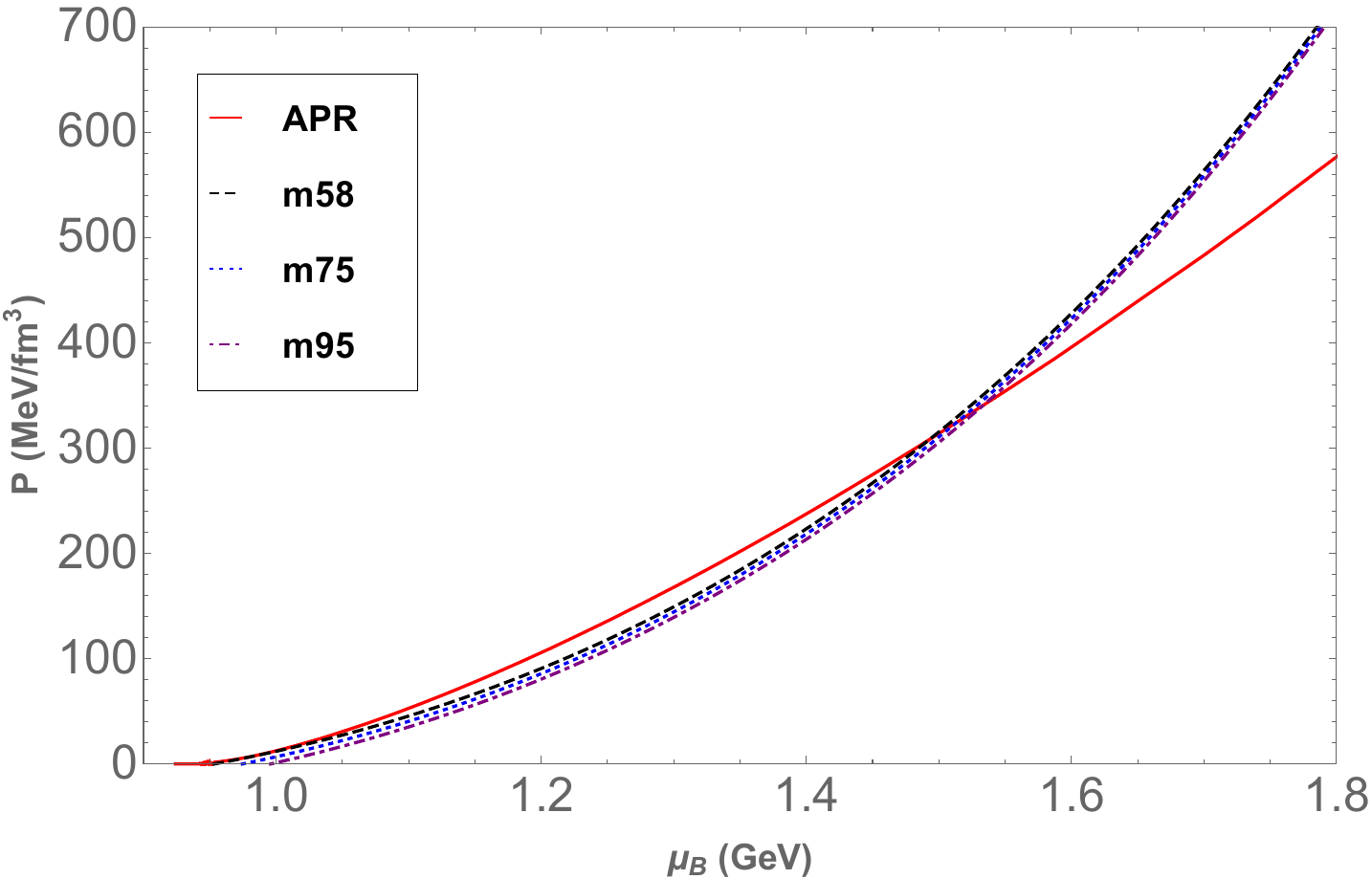}}
\subfloat{\includegraphics[width=9cm,height=6cm]{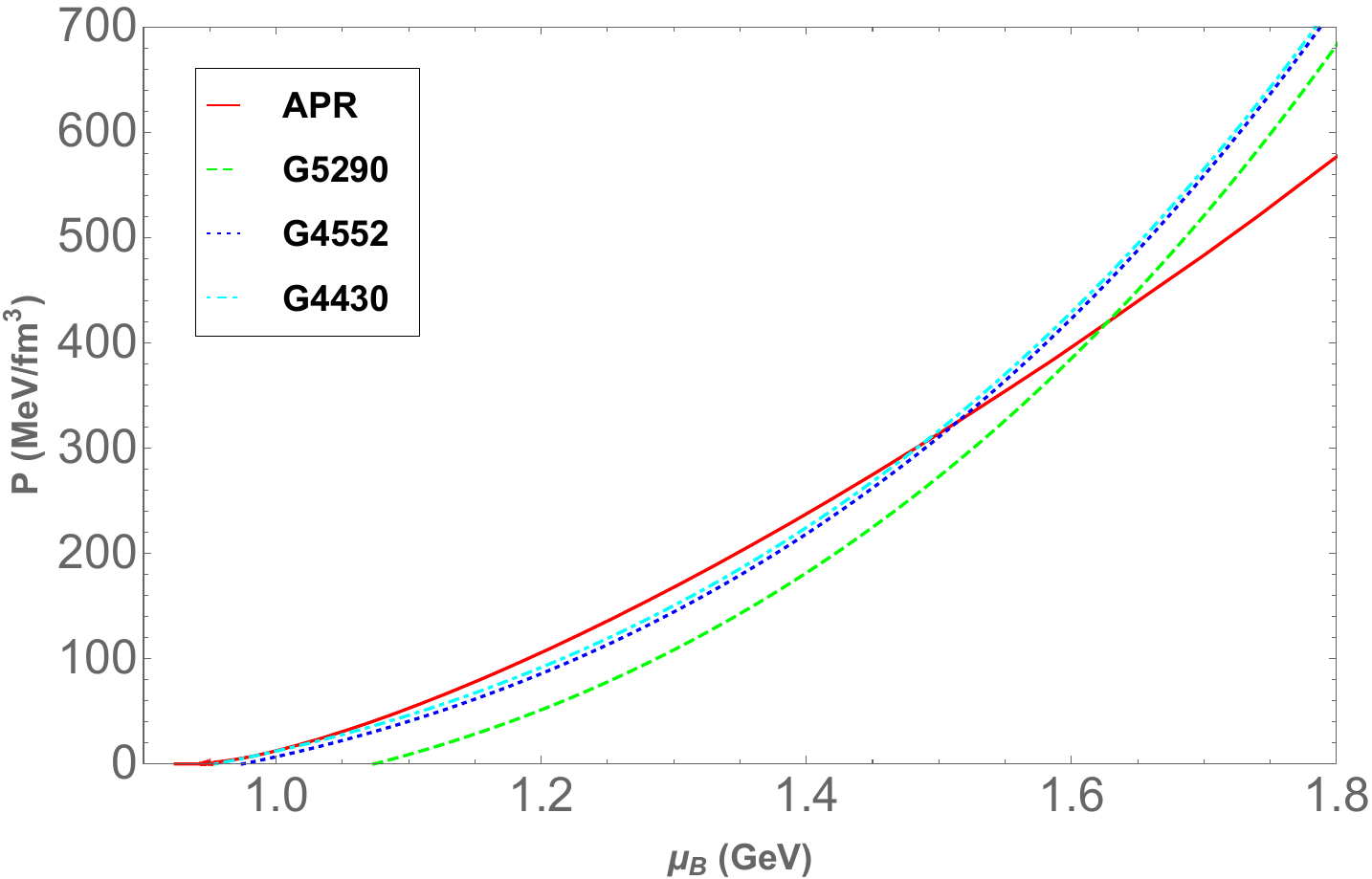}}
\caption{The pressure versus baryon chemical potential
($\mu_{\rm B}$) for $T_{\rm pc}=181$ MeV
(left panel) and $m=7.5$ MeV (right panel), respectively.
The pressure of the hadronic matter described by BSK21
is also presented for a comparison. Here, m58, m75, m95
refer to $m=5.8$, 7.5, 9.5 MeV, respectively,
and G5290, G4552, G4430 refer to
$G_1=5.290$, 4.552, 4.430 GeV$^{-2}$, respectively.}
\label{Figure6}
\label{groupa}
\end{figure*}

In Fig.~\ref{Figure6}, the influences of $m$ and $G_1$
on the pressure of the quark matter are presented in the
left panel and right panel with
$T_{\rm pc}=181$ MeV and $m=7.5$ MeV, respectively.
We can see that after considering the constraint of lattice
simulation results on $T_{\rm pc}$, the EOSs of quark matter
under different parameter sets are very close to each other.
It could also be seen that the deconfinement phase transition
occurs at about $\mu_B\sim1.5$ GeV. The left panel
shows the EOSs corresponding to $T_{\rm pc}=181$ MeV and
$m=5.8$, 7.5, 9.5 MeV, respectively.
Similarly, the right panel shows the EOSs corresponding to
$m=7.5$ MeV and $G_1=5.290$, 4.552, 4.430 GeV$^{-2}$,
respectively. The EOSs with $G_1=4.430$ and 4.552 GeV$^{-2}$
are very close to each other and their corresponding
$T_{\rm pc}$ values are in accordance with
the lattice simulation results. However,
both of these two EOSs are quite different from that
of $G_1=5.290$ GeV$^{-2}$, which corresponds to the case of
normal NJL model.

\begin{figure*}
\centering
\subfloat{\includegraphics[width=9cm,height=6cm]{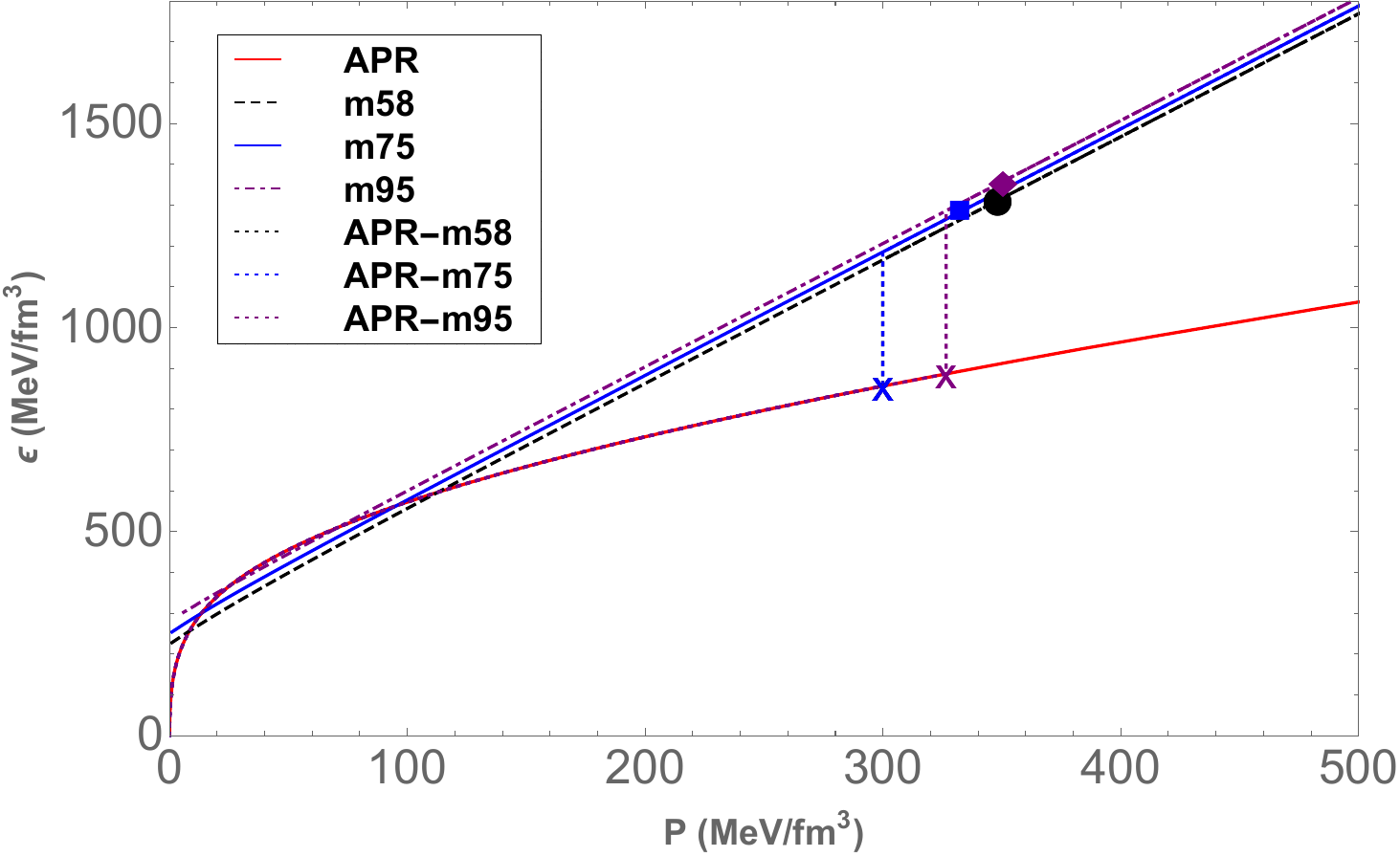}}
\subfloat{\includegraphics[width=9cm,height=6cm]{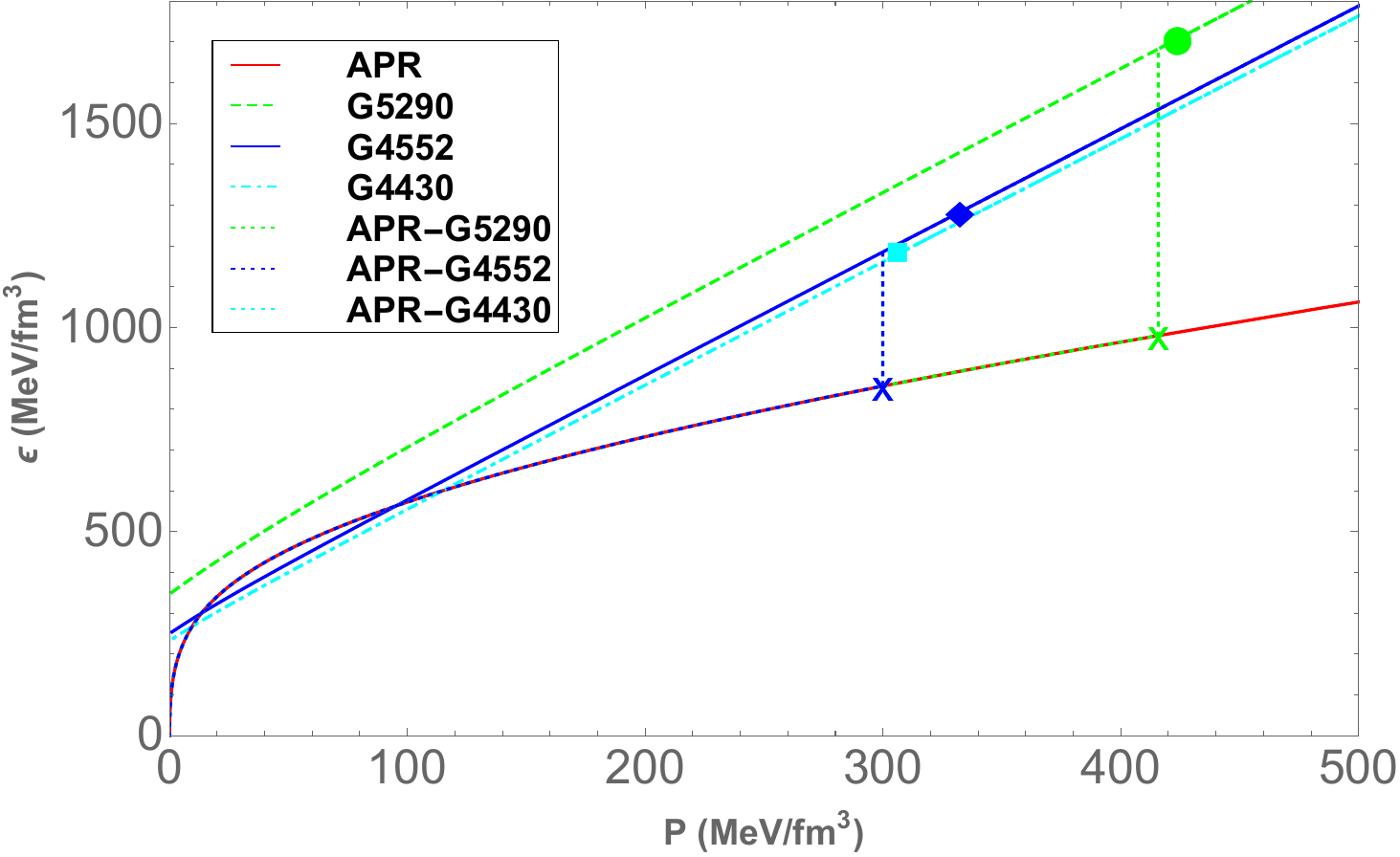}}
\caption{The $\epsilon-P$ relation of the quark matter and
hybrid EOS when $T_{\rm pc}=181$ MeV (left panel)
and $m=7.5$ MeV (right panel), respectively.
The $\epsilon-P$ relation of the hadronic matter
described by APR is shown by the red line.
Here, m58, m75, m95 (APR-m58, APR-m75, APR-m95)
refer to the quark matter (hybrid EOS) with
$m=5.8$, 7.5, 9.5 MeV, respectively, and
G5290, G4552, G4430 (APR-G5290, APR-G4552, APR-G4430)
refer to quark matter (hybrid EOS) with
$G_1=5.290$, 4.552, 4.430 GeV$^{-2}$, respectively.}
\label{Figure7}
\label{groupa2}
\end{figure*}

In Fig.~\ref{Figure7}, we present the $\epsilon-P$ relations of
the hadronic matter, the quark matter and hybrid EOSs with the
Maxwell construction. Each point marked with ``x'' represents
the critical point of the corresponding first-order
phase transition, which is denoted as ($P_{\rm pt}$,
$\epsilon_{\rm pt}$) hereafter. The other marked point
on each hybrid EOS refers to the center of
the most massive hybrid star. Here we denote it as
($P_{\rm c}$, $\epsilon_{\rm c}$). We can see that
according to the constraint of lattice simulation results
on $T_{\rm pc}$, $\epsilon_{\rm c}$ ($\epsilon_{\rm pt}$)
of the hybrid EOSs only varies slightly among
different parameter sets. Generally, a larger $G_1$ will
lead to a larger value $P_{\rm c}$ and $\epsilon_{\rm c}$.
The corresponding ($P_{\rm pt}$, $\epsilon_{\rm pt}$)
and ($P_{\rm c}$, $\epsilon_{\rm c}$) points of five
representative hybrid EOSs are listed in Table~\ref{Table2}.

\begin{table}
\centering
\caption{The corresponding ($P_{\rm pt}$, $\epsilon_{\rm pt}$)
and ($P_{\rm c}$, $\epsilon_{\rm c}$) points of five
representative hybrid EOSs.}
\label{Table2}
\begin{tabular}{p{1.2cm} p{1.3cm} p{2.3cm}p{2.3cm}}
    \hline\hline
    $\quad\, m$&$\quad\,\, G_1$&$\,\,\,\,\,(P_{pt},
    \epsilon_{pt})$&$\,\,\,\,\,\,\,(P_{\rm c}, \epsilon_{\rm
    c})$\\
    $\,\,[{\rm MeV}]$&$[{\rm GeV}^{-2}]$&$\,[{\rm MeV}
    \cdot{\rm fm}^{-3}]$&$\,[{\rm MeV}\cdot{\rm fm}^{-3}]$\\
    \hline
    $\quad$5.8&$\,\,\,\,3.485$&$(299.9, 856.5)$&
    $(348.6, 1312.3)$\\
    \hline
    $\,$&$\,\,\,\,4.430$&$(299.9, 856.5)$&
    $(306.2, 1179.6)$\\
    $\quad$7.5&$\,\,\,\,4.552$&$(299.9, 856.5)$&
    $(332.5, 1283.2)$\\
    $\,$&$\,\,\,\,5.290$&$(415.8, 980.1)$&
    $(424.1, 1707.7)$\\
    \hline
    $\quad$9.5&$\,\,\,\,5.572$&$(326.6, 886.3)$&
    $(350.5, 1358.2)$\\
    \hline\hline
\end{tabular}
\end{table}

\begin{figure*}
\centering
\subfloat{\includegraphics[width=9cm,height=6cm]{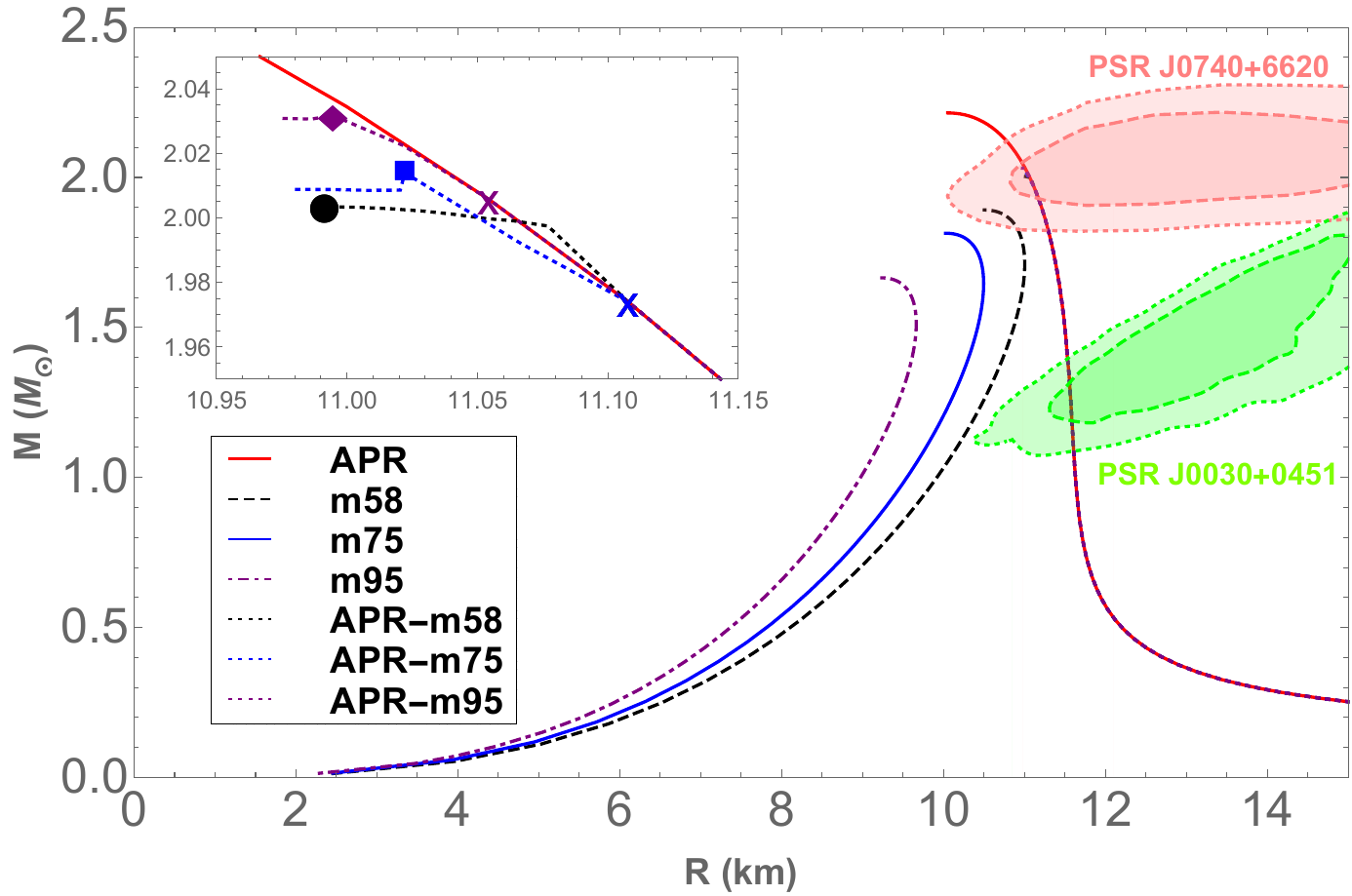}}
\subfloat{\includegraphics[width=9cm,height=6cm]{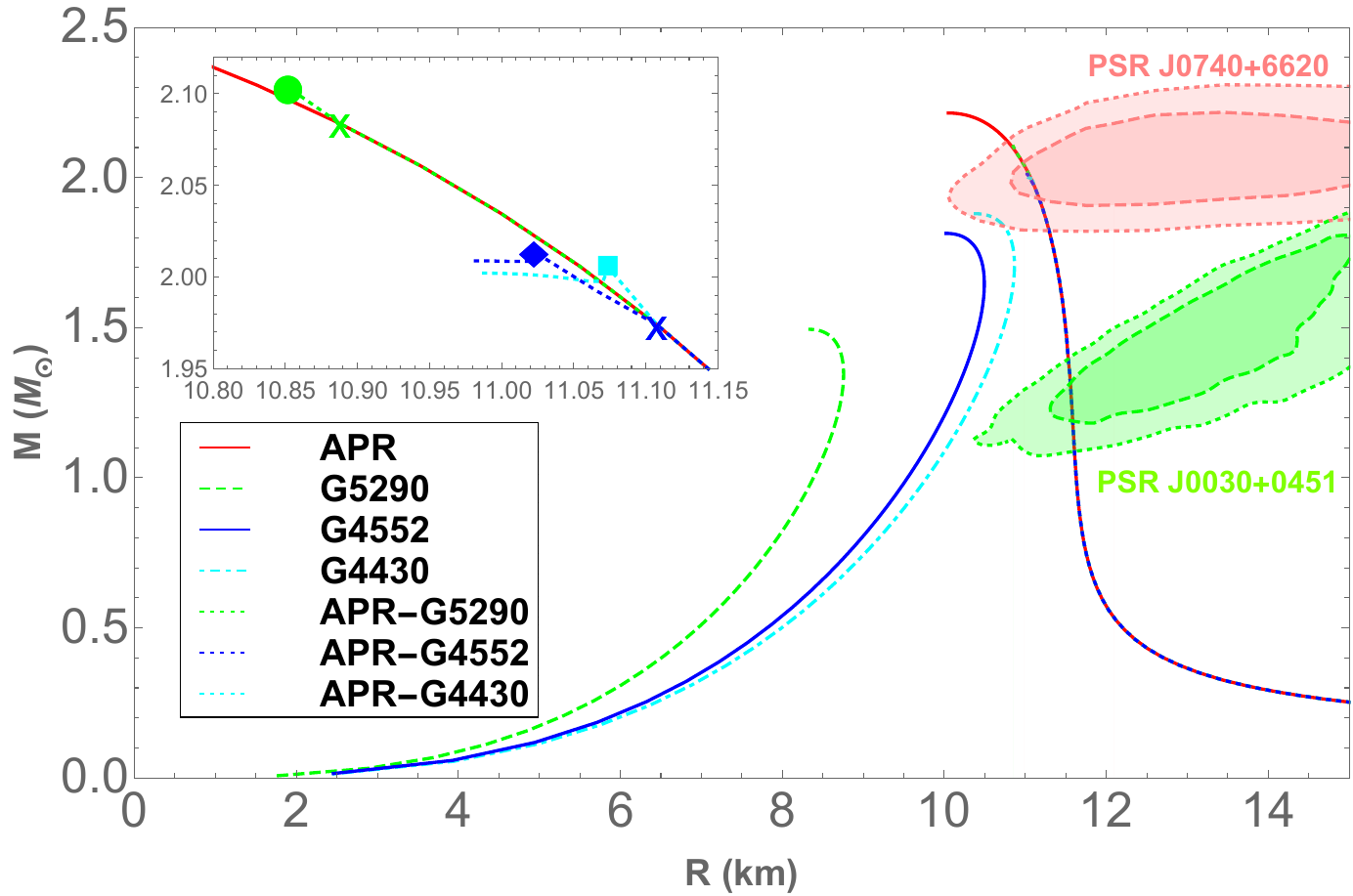}}
\caption{The $M-R$ relations of quark stars and
hybrid stars for $T_{\rm pc}=181$ MeV (left panel)
and $m=7.5$ MeV (right panel), respectively. Line styles
are the same as in Fig.~\ref{Figure7}.}
\label{Figure8}
\label{groupa3}
\end{figure*}

\begin{figure*}
\centering
\subfloat{\includegraphics[width=9cm,height=6cm]{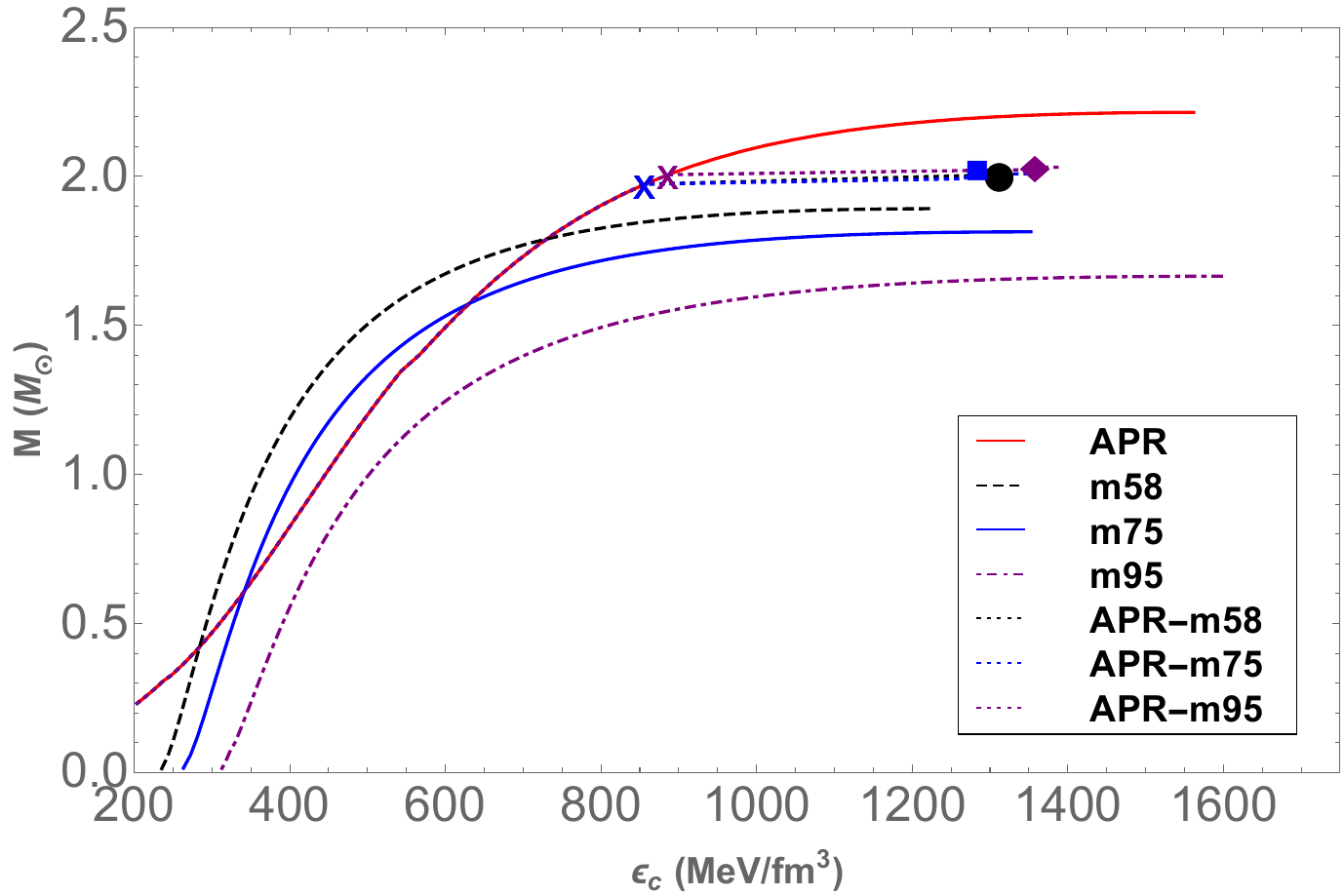}}
\subfloat{\includegraphics[width=9cm,height=6cm]{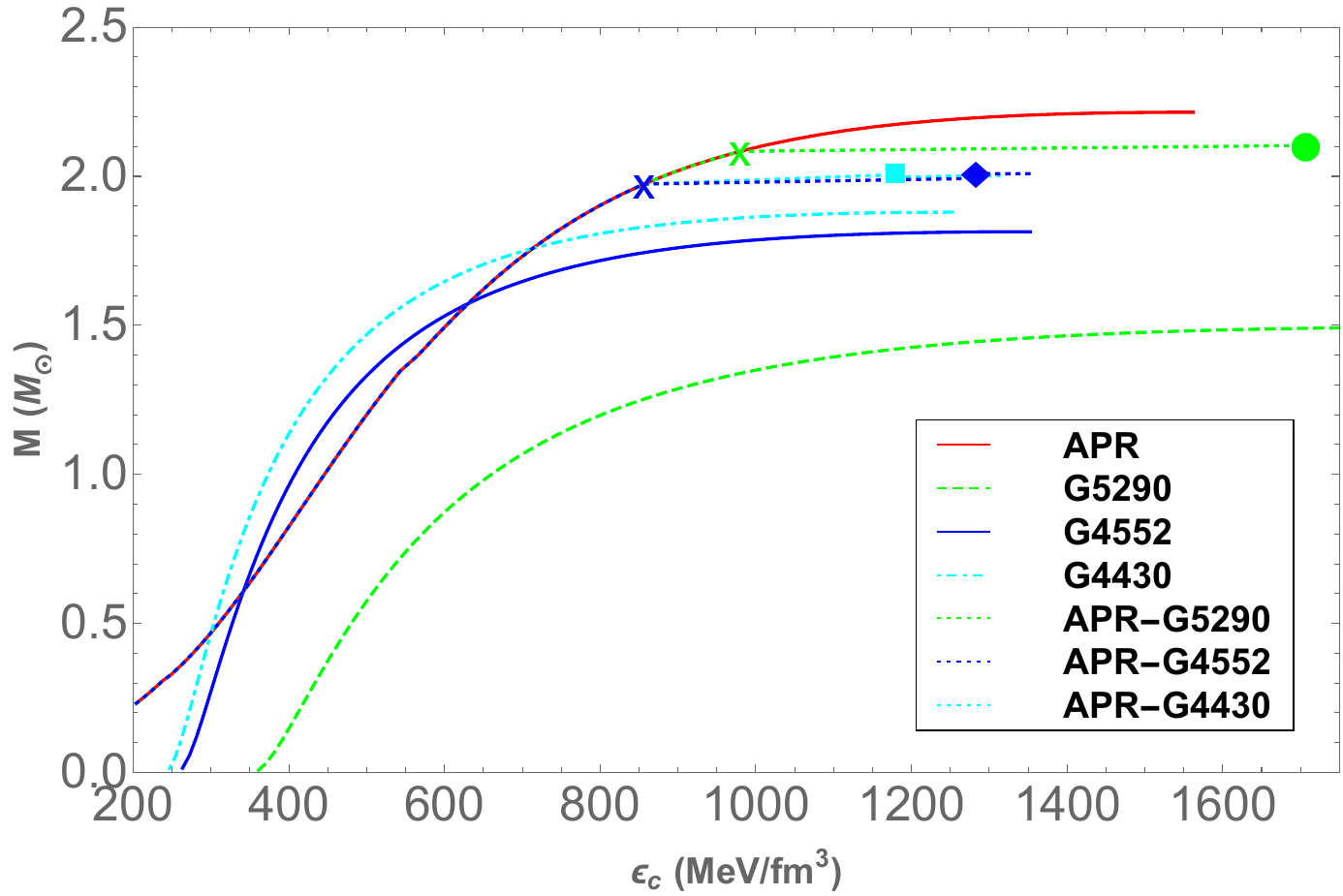}}
\caption{The $M-\epsilon_{\rm c}$ relations of quark stars
and hybrid stars for $T_{\rm pc}=181$ MeV (left panel) and
$m=7.5$ MeV (right panel), respectively. Line styles are
the same as in Fig.~\ref{Figure7}.}
\label{Figure9}
\label{groupa4}
\end{figure*}

Once the EOS is determined, we can solve the
Tolman-Oppenheimer-Volkoff (TOV) equation numerically
to get the $M-R$ and mass-central energy density
($M-\epsilon_{\rm c}$) relations. In Fig.~\ref{Figure8} and
Fig.~\ref{Figure9}, to study the influence of $m$ and $G_1$
on the $M-R$ and $M-\epsilon_{\rm c}$
relations, the corresponding results for
$T_{\rm pc}=181$ MeV and $m=7.5$ MeV
are shown in the left panel and right panel, respectively. In
Fig.~\ref{Figure8}, the most massive quark star for
$m=5.8$, 7.5, 9.5 MeV ($G_1=5.290$, 4.552, 4.430 GeV$^{-2}$)
is about 1.892 $M_{\odot}$ (1.879 $M_{\odot}$),
not satisfying the 2 $M_{\odot}$ constraint.
In addition, the quark stars cannot fulfill the recent
$M-R$ constraint from the NICER measurement of PSR
J0030+0451~\citep{Miller_2019}, although the other
constraint from PSR J0740+6620~\citep{Miller_2021} is
marginally fulfilled by the quark EOSs
with the parameter sets of $(m, G_1)=(5.8, 3.485)$
and $(7.5, 4.430)$. However, the hybrid EOSs obtained with the
Maxwell construction approach in this work can produce
hybrid stars in consistent with these
astronomical observations, although their quark matter cores
are relatively small (about 0.03 $M_{\odot}$).

The $M-\epsilon_{\rm c}$ relations are shown
in Fig.~\ref{Figure9}. We can find that
for stable neutron stars (whether they are hadron stars,
quark stars, or hybrid stars), a larger $\epsilon_{\rm c}$
corresponds to a more massive star.
The $\epsilon_{\rm pt}$ ($\epsilon_{\rm c}$) of
the hybrid stars whose corresponding hybrid EOSs satisfy the
constraint of the lattice simulation results on $T_{\rm pc}$
is in a range of  $856-886$ ($1180-1358$)
MeV/fm$^3$, which can also be seen in Table~\ref{Table2}.
However, for the normal NJL model with $m=7.5$ MeV,
the corresponding $\epsilon_{\rm pt}$ ($\epsilon_{\rm c}$)
of hybrid stars is about 980 (1700) MeV/fm$^3$,
quite different from the results of the modified NJL model.

We have also calculated the tidal deformability of
hadron stars, quark stars and hybrid stars in this work,
which is defined as~\citep{PhysRevD.81.123016},
\begin{equation}\label{TD}
  \Lambda=\frac{2}{3}k_2R^5.
\end{equation}
Here $k_2$ is the dimensionless tidal Love number for $l=2$,
which can be calculated by
\begin{eqnarray}
  &k_2&=\frac{8C^5}{5}(1-2C)^2[2+2C(y-1)-y]\nonumber\\
  &\times&\{2C[6-3y+3C(5y-8)]\nonumber\\
  &+&4C^3[13-11y+C(3y-2)+2C^2(1+y)]\nonumber\\
  &+&3(1-2C)^2[2+2C(y-1)-y]ln(1-2C)\}^{-1},\label{tln}
\end{eqnarray}
where $C=M/R$ refers to the compactness of the star, and $y$
is defined as
\begin{equation}\label{parametery}
  y=R\beta(R)/H(R)-4\pi R^3\epsilon_0/M,
\end{equation}
where $\epsilon_0$ is the energy density at the surface of
the star. The dimensionless parameter $y$ can be calculated
by solving the differential equations
~\citep{PhysRevD.81.123016},
\begin{eqnarray}
  \frac{dH}{dr} &=& \beta,\nonumber\\
  \frac{d\beta}{dr} &=&
  2(1-2\frac{m_r}{r})^{-1}H\{-2\pi[5\epsilon+9P+f(\epsilon+P)]\nonumber\\
  &+&\frac{3}{r^2}+2(1-2\frac{m_r}{r})^{-1}(\frac{m_r}{r^2}
  +4\pi rP)^2\}\nonumber\\
  &+&\frac{2\beta}{r}(1-2\frac{m_r}{r})^{-1}\{\frac{m_r}{r}
  +2\pi r^2(\epsilon-P)-1\},\,\label{HbetaEq}
\end{eqnarray}
where $H(r)$ is the metric function,
and $f={\rm d}\epsilon/{\rm d}P$.

\begin{figure}
\includegraphics[width=0.47\textwidth]{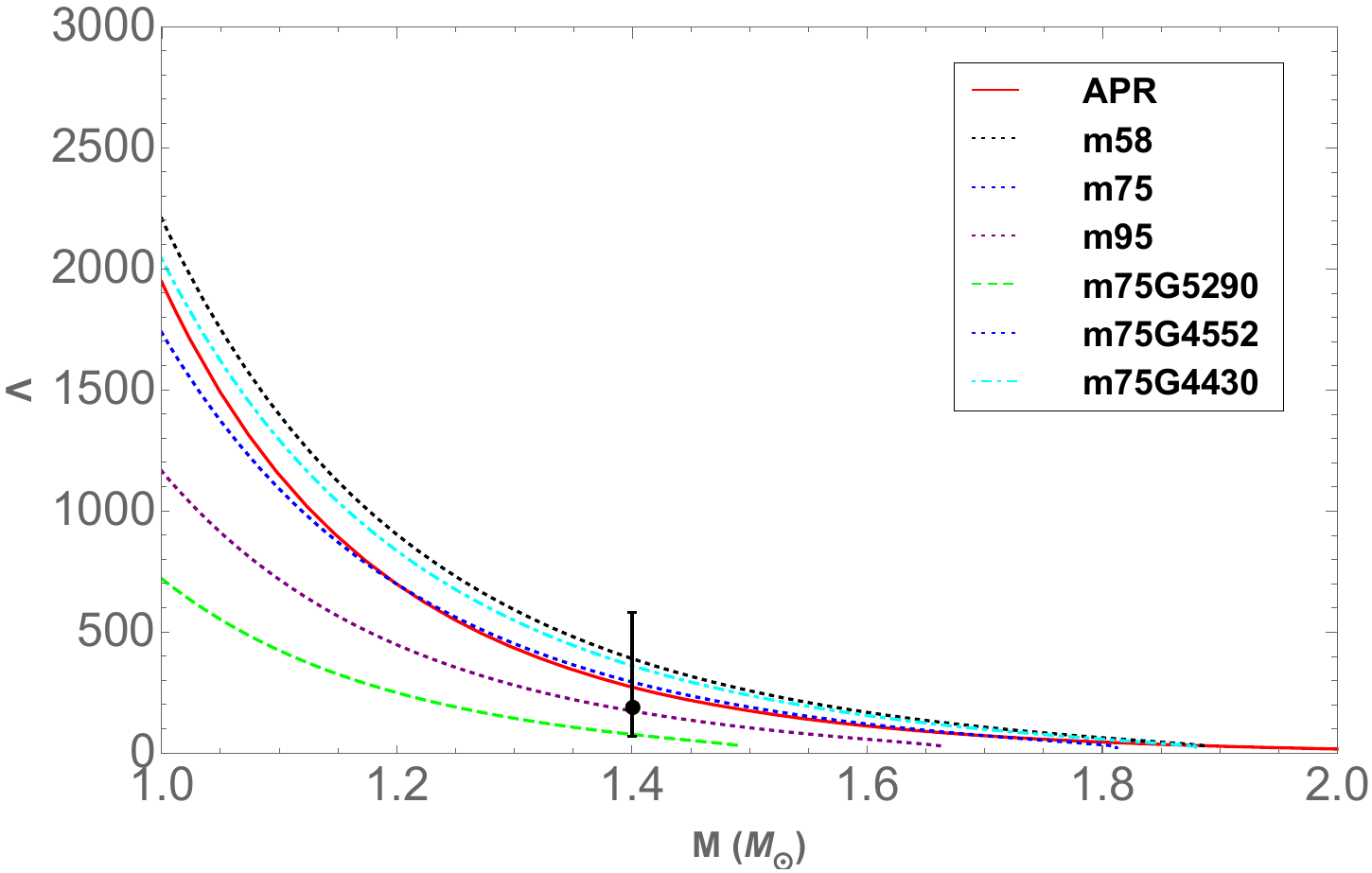}
\caption{The $\Lambda-M$ relation of quark stars and
hadron stars. Here, m58, m75, m95 refer to quark stars
with $T_{\rm pc}=181$ MeV and $m=5.8$, 7.5, 9.5 MeV,
respectively, and m75G5290, m75G4552, m75G4430 refer to
quark stars with $m=7.5$ MeV and
$G_1=5.290$, 4.552, 4.430 GeV$^{-2}$, respectively.
The observational constraint from GW170817,
$\Lambda(1.4 M_{\odot})=190_{-120}^{+390}$
~\citep{PhysRevLett.121.161101} is also plotted
for comparison.} \label{Figure10}
\end{figure}

The $\Lambda-M$ relation is shown in
Fig.~\ref{Figure10}. We can see that for quark stars
and hadron stars described by the modified NJL model and APR
hadronic model, the corresponding values of
$\Lambda(1.4 M_{\odot})$ satisfy the constraint from GW170817,
i.e., $\Lambda(1.4 M_{\odot})=190_{-120}^{+390}$
~\citep{PhysRevLett.121.161101}. For stable hybrid stars
whose maximum masses are higher than 2 $M_{\odot}$
in this work, the corresponding hybrid EOSs demonstrate that
neutron stars with the masses lower than 1.974 $M_{\odot}$
are still hadron stars and the quark matter cores
do not exist inside them. Therefore, the $\Lambda-M$ relations
from these hybrid EOSs are the same with that of hadron stars
when $M\leq1.974$ $M_{\odot}$, and we do not show $\Lambda-M$
relations of hybrid stars in Fig.~\ref{Figure10}. In other
words, according to the hybrid EOSs constrained in this work,
the BNS in GW170817 whose masses are estimated to be
$1.17-1.36$ and $1.36-1.60$
$M_{\odot}$~\citep{PhysRevLett.119.161101},
respectively, should both be hadron stars.

For the sake of completeness, the $M-R$ properties of
hybrid stars constructed by five representative hybrid EOSs
are presented in Table~\ref{Table3},
where ($R_{pt}$, $M_{pt}$) is related to the hadron-quark
phase transition point, referring to the radius and mass
of the most massive hadron star constructed by the hybrid EOS,
and ($R_{\rm max}$, $M_{\rm max}$) is the radius and mass of
the most massive hybrid star with a quark matter core.
We can see that the masses of quark matter cores
are in a range of $0.026-0.04$ $M_{\odot}$
in the framework of the modified NJL model.
For a fixed $T_{\rm pc}$ ($m$), a larger $m$ ($G_1$) leads to
a larger $M_{\rm max}$. Due to the constraint of $T_{\rm pc}$
from the lattice simulation results, $M_{\rm max}$ is about
0.1 $M_{\odot}$ lower than that of normal NJL model,
while the quark matter core is about 0.015 $M_{\odot}$ heavier
than that of normal NJL model.

\begin{table}
\centering
\caption{The corresponding ($R_{pt}$, $M_{pt}$) and
($R_{\rm max}$, $M_{\rm max}$) points of five representative
hybrid EOSs.}
\label{Table3}
\begin{tabular}{p{1.2cm} p{1.3cm} p{2.3cm}p{2.3cm}}
    \hline\hline
    $\quad\, m$&$\quad\,\, G_1$&$\,\,\,\,(R_{pt},
    M_{pt})$&$(R_{\rm max}, M_{\rm max})$\\
    $\,\,[{\rm MeV}]$&$[{\rm GeV}^{-2}]$&$\,\,\,\,\,[km,
    M_{\odot}]$&$\,\,\,\,\,\,[km, M_{\odot}]$\\
    \hline
    $\quad$5.8&$\,\,\,\,3.485$&$(11.11, 1.974)$&$\,(10.99,
    2.003)$\\
    \hline
    $\,$&$\,\,\,\,4.430$&$(11.11, 1.974)$&$\,(11.07, 2.005)$\\
    $\quad$7.5&$\,\,\,\,4.552$&$(11.11, 1.974)$&$\,(11.02,
    2.014)$\\
    $\,$&$\,\,\,\,5.290$&$(10.89, 2.084)$&$\,(10.85, 2.103)$\\
    \hline
    $\quad$9.5&$\,\,\,\,5.572$&$(11.05, 2.006)$&$\,(10.99,
    2.032)$\\
    \hline\hline
\end{tabular}
\end{table}

\section{Summary}
\label{three}

In this study, the modified 2-flavor NJL model and
the APR EOS with $A18+\delta\nu+UIX^{\ast}$ interaction
are introduced to investigate the nonstrange quark matter and
hadronic matter in hybrid stars in light of a hypothesis that
the quark matter may not be strange. To construct hybrid EOSs,
the first-order hadron-quark phase transition and
the corresponding Maxwell construction are considered.

It is noted that the modification of the coupling constant $G$
in the normal NJL model is helpful, because it is not only
consistent with the QCD requirement in essence, but also
in agreement with the lattice simulation results of
$T_{\rm c}$. In the 2-flavor case,
when $T_{\rm c}=173\pm8$ MeV, the parameter space of $G_1$
can be constrained for a given current quark mass $m$,
and the corresponding range of $G_2$ can also be determined.
The results are shown in Table~\ref{Table1},
implying that the normal 2-flavor NJL model with the four-quark
scalar interaction (corresponding to the case of $G_2=0$ in our
modified NJL model) is inconsistent with the lattice simulation
results. For hybrid EOSs, the influence of $m$ and $G_1$
is very small when the constraint of $T_{\rm c}=173\pm8$ MeV
is taken into account. The hybrid EOSs derived from
the modified NJL model are quite different from that of
normal NJL model.

Considering astronomical observations and the stability of
hybrid stars, the parameter $m$ is constrained to be
$5.8-9.5$ MeV, and the parameter space of $G_1$ can also
be determined. The quark EOSs constructed with the modified
NJL model in this work is soft, and thus cannot satisfy
the 2 $M_{\odot}$ constraint of neutron stars and
some $M-R$ constraints from NICER missions. It is noted that
in some previous studies, the quark matter cores may not exist
in compact stars under the Maxwell construction
~\citep{ozel2006soft,PhysRevLett.117.032501,PhysRevD.107.103009},
or the maximum mass of quark matter cores may be larger than
0.6 $M_{\odot}$ and the BNS in GW170817 can be hybrid stars
~\citep{ayriyan2021bayesian,universe6060081,universe5020061,PhysRevC.105.035802}.
However, with the modified 2-flavor NJL model in this work,
the hybrid EOSs with first-order hadron-quark transitions
are still in agreement with current neutron star
astronomical observations, and pure nonstrange quark matter
cores can exist in hybrid stars, possessing a relatively
small mass of $0.026-0.04$ $M_{\odot}$. Due to the
constraint of $T_{\rm c}=173\pm8$ MeV, $M_{\rm max}$ of
hybrid stars will be about 0.1 $M_{\odot}$ lower than that of
normal NJL model. According to the hybrid EOSs constrained
in this work, the BNS in GW170817 whose masses are estimated
to be $1.17-1.36$ and $1.36-1.60$
$M_{\odot}$~\citep{PhysRevLett.119.161101}, respectively,
may be hadron stars.

\begin{acknowledgments}

We thank the anonymous referee for useful suggestions
that led to an overall improvement of the presentation.
This study is supported in part by the
National Key Program for Science and Technology Research
Development (2023YFB3002500),
the National Natural Science Foundation of China
(under Grants No. 12005192, No. 12075213, and No. 12233002),
the Project funded by China Postdoctoral Science Foundation
(No. 2020M672255, No. 2020TQ0287),
National SKA Program of China No. 2020SKA0120300,
the National Key R$\&$D Program of China (2021YFA0718500),
the Natural Science Foundation of Henan Province of China
(No. 242300421375),
the Natural Science Foundation for Distinguished
Young Scholars of Henan Province under grant number
242300421046,
the start-up funding from Zhengzhou University.
Y.F.H also acknowledges the support from
the Xinjiang Tianchi Program.

\end{acknowledgments}

\bibliographystyle{aasjournal}
\bibliography{reference}

\end{document}